\documentclass[a4paper,fleqn,usenatbib]{mnras}

\usepackage{newtxtext,newtxmath}

\usepackage[T1]{fontenc}
\usepackage{ae,aecompl}

\usepackage{graphicx}	
\usepackage{amsmath}	
\usepackage{amssymb}	

\usepackage{epsfig}
\usepackage{multirow}
\usepackage{float}
\def\mnras{MNRAS}
\def\apj{ApJ}
\def\aj{AJ}
\def\apjl{ApJL}
\def\apjs{ApJS}
\def\aap{A\& A}
\def\pasp{PASP}
\def\pasj{PASJ}
\def\nat{Nature}

\def\xmm{{\it XMM-Newton}}

\def\rosat{{\it ROSAT}}
\def\rxj04{RX J0439.6-5311}
\def\1h07{1H 0707-495}
\def\pg12{PG 1244+026}

\title[RX J0439.6-5311. II. Global Structure of the Accretion Flow]{Super-Eddington QSO RX J0439.6-5311. II. Multi-wavelength Constraints on the Global Structure of the Accretion Flow}
\author[C. Jin, et al.]{
Chichuan Jin$^{1}$\thanks{E-mail: chichuan@mpe.mpg.de},
Chris Done$^{2}$,
Martin Ward$^{2}$,
Emma Gardner$^{1}$
\\
$^{1}$Max-Planck-Institut f\"{u}r Extraterrestrische Physik, Giessenbachstrasse, D-85748 Garching, Germany\\
$^{2}$Centre for Extragalactic Astronomy, Department of Physics, University of Durham, South Road, Durham DH1 3LE, UK\\
}

\date{prepared for MNRAS}

\pubyear{2017}

\begin{document}
\label{firstpage}
\pagerange{\pageref{firstpage}--\pageref{lastpage}}
\maketitle

\begin{abstract}
  We present a detailed multi-wavelength study of an unobscured,
 highly super-Eddington Type-1 QSO \rxj04. We combine the latest
 \xmm\ observation with all archival data from infrared to hard
 X-rays. The optical spectrum is very similar to that of \1h07\
 in having extremely weak [O {\sc iii}] and strong Fe {\sc ii} emission lines, although the
 black hole mass is probably slightly higher at $5-10
 \times10^{6}~\rm M_{\odot}$. The broadband SED is uniquely well
 defined due to the extremely low Galactic and intrinsic absorption, so 
the bolometric luminosity is tightly constrained. 
 The optical/UV accretion disc continuum is seen
 down to 900 \AA, showing that there is a  standard thin disc
 structure down to $R \ge$ 190-380 $R_{\rm g}$  and determining the mass accretion rate through the outer disc.
This predicts a much higher bolometric luminosity than observed, indicating that there must be strong wind and/or advective energy losses from the inner disc, as expected for a highly super-Eddington accretion flow.
 Significant outflows are detected in both the NLR and BLR emission lines, confirming
the presence of a wind. We propose a global picture for the structure of a super-Eddington
accretion flow where the inner disc puffs up, shielding much of the potential NLR material,
and show how inclination angle with respect to this and the wind can explain very different
X-ray properties of RX J0439.6-5311 and 1H 0707-495. Therefore, this source provides strong
supporting evidence that `simple' and `complex' super-Eddington NLS1s can be unified within
the same accretion flow scenario but with different inclination angles. We also propose that
these extreme NLS1s could be the low-redshift analogs of weak emission-line quasars (WLQs).

\end{abstract}




\section{Introduction}
Active Galactic Nuclei (AGN) are powered by the accretion of material
onto a super-massive black hole (SMBH). Hence their emission should be
determined mainly by the fundamental parameters of mass and spin of
the black hole together with mass accretion rate, while inclination
angle can also affect the observed properties.  The current zoo of AGN
subtypes should then map to these parameters. Boroson (2002) show how
the optical spectra of different types of unobscured AGN can be
decomposed in terms of principle components (PC), and that the
most important (PC1: weak [O {\sc iii}], strong Fe {\sc ii} and narrow H$\beta$;
PC2: weak He {\sc ii}) corresponded to increasing luminosity relative to
Eddington, $L/L_{\rm Edd}$ and total luminosity, respectively. This
identifies the optical class of Narrow Line Seyfert 1 (NLS1) galaxies
(Osterbrock \& Pogge 1985) as powered by high $L/L_{\rm Edd}$ accretion
onto a low mass ($\lesssim 10^7~\rm M_\odot$) black hole.

NLS1s also typically show strong soft X-ray emission (below 2~keV:
Boller, Brandt \& Fink 1996), and a steeper 2-10~keV X-ray spectrum
than broad line Seyfert 1s (Brandt, Mathur \& Elvis
1997). Extrapolating their 2-10~keV spectra down below 2~keV reveals a
strong soft X-ray excess above the prediction of a single power law.
This excess is interpreted as some combination of either 
relativistically smeared, ionised reflection from the inner disc
(e.g. Miniutti \& Fabian 2004; Ross \& Fabian 2005; Fabian \& Miniutti
2005; Crummy et al. 2006; Fabian et al. 2013), Comptonisation from an
additional electron population at $\sim0.2$ keV (e.g. Laor et
al. 1997; Magdziarz et al. 1998; Gierli\'{n}ski \& Done 2004; Done et
al. 2012 hereafter D12), and intrinsic emission from the accretion
disc itself (D12; Jin et al. 2013; Chiang et al. 2015).

The 2-10~keV X-rays are variable on short timescales, as expected for
a low mass black hole (e.g. Ponti et al. 2012). However, some fraction
of NLS1s show extreme variability amplitudes, with strong spectral
variability seen in low flux states, where their 2-10~keV spectra are
much flatter, with extremely strong features seen around the iron K
shell energy at 6-8~keV (Gallo 2006). The spectra of these `complex'
NLS1s (as opposed to the `simple' NLS1s described above: Gallo 2006)
have been variously interpreted as being dominated by highly
relativistic ionised reflection from the very inner disc (Fabian et
al. 2013); and/or absorption and scattering by material in a wind
(e.g. Turner et al. 2007; Miller et al. 2007; Sim et al. 2010; Tatum
et al. 2012; Gardner \& Done 2015; Hagino et al. 2016). The best known
examples of this complex NLS1 class are \1h07\ and IRAS13224-3809
(Leighly \& Moore 2004; Ponti et al. 2010;  Chiang et al. 2015; Parker et al. 2017), while well known `simple'
ones are \pg12\, RE~J1034+324 (the QPO AGN: Gierli\'{n}ski et al. 2008; Jin
et al. 2013), Ton S180 and Akn 564 (Gallo 2006).

Understanding the accretion flow in NLS1 requires breaking the spectral degeneracies
which underlie these different physical models. One way to do this is to extend the
bandpass to study the broader Spectral Energy Distribution (SED), to set the X-ray
spectrum in a multi-wavelength context. Optical/UV data in particular
can constrain the two key parameters of both mass and mass accretion
rate. Done \& Jin (2016) (hereafter: DJ16) compared two well studied
NLS1s, \pg12\ as an archetypal `simple' NLS1 and \1h07\ as a
archetypal `complex' NLS1. Both sources have similarly low black hole
masses (similarly narrow H$\beta$ line, similar optical continuum
luminosity) and similar absolute mass accretion rates, $\dot{M}$,
(determined from the optical continuum: $F_{\rm opt}\propto \cos i
~(M\dot{M})^{2/3}$, Davis \& Laor 2011). Both are then similarly at
Eddington or above even for the highest possible mass, zero spin and
low inclination angle (DJ16). Low mass, high spin and high inclination
(as derived for \1h07\ from reverberation mapping assuming an inner disc
reflection dominated spectrum e.g. Fabian et al. 2009; Zoghbi et al.
2010; Kara et al. 2013) mean that the mass accretion rate is 
extremely super-Eddington, implying $L\ge 150~L_{\rm Edd}$ for \1h07\ (DJ16).

A highly super-Eddington accretion flow is very likely to power a wind
(Ohsuga \& Mineshige 2011; Jiang, Davis \& Stone 2014; Takeuchi, Ohsuga \& Mineshige 2014).
This energy loss can explain the apparent discrepancy
between the observed X-ray spectrum and the much higher X-ray flux
predicted by a standard accretion disc spectrum in both \pg12\ and
\1h07\ (DJ16), as well as in the intermediate-mass black hole RX
J1140.1+0307 (Jin, Done \& Ward 2016).  In the presence of a clumpy
disc wind, the difference between `simple' and `complex' NLS1s can be
explained as due to different inclination angles (DJ16). Gardner \&
Done (2015) showed that absorption caused by gas clumps above the disc
with different viewing angles can reproduce the observed time lag
transition from the 200 s hard X-ray reverberation lag in \pg12\ to
the 30 s lag in \1h07. Hagino et al. (2016) showed that absorption by
wind clumps can also reproduce the X-ray spectra of \1h07\ above 2
keV, including the broad iron K$\alpha$ feature, without requiring any
extreme relativistic smearing or super-solar iron abundance (see also
the recent detection of blueshifted absorption from a wind in the
complex NLS1 IRAS13224+3809: Parker et al. 2017). These studies
support the view that it is the inclination angle which is the key
parameter separating the two X-ray classes of NLS1s.

However, there are some other differences between \pg12\ and \1h07\
which cannot be explained as being solely due to different inclination
angles. For example, forbidden lines such as the [O {\sc
  iii}]$\lambda$4959/5007 doublets are much weaker in \1h07\ than in
\pg12. These make \1h07\ appear more extreme on PC1 (Boroson 2002),
implying higher $L/L_{\rm Edd}$ than in \pg12. Given that they have the
same optical flux, the $\cos i$ dependence of the accretion disc
emission means that this does imply a higher $L/L_{\rm Edd}$ for \1h07, even
if it has the same mass and spin as \pg12. For super-Eddington fluxes,
the inner disc is expected to puff up, shielding more and more of the
narrow line region (NLR) clouds for higher $L/L_{Edd}$, hence
reducing the [O {\sc iii}] flux (e.g. Leighly 2004; Luo et al. 2015).

In any case, the inclination angle scenario predicts that there should
be super-Eddington NLS1s with X-ray spectra similar to `simple' NLS1s
like \pg12, but weak NLR lines similar to \1h07. Recently, Jin,
Done \& Ward (2017) (hereafter: Paper-I) presented a detailed X-ray
analysis of such an AGN, namely \rxj04, which has the smallest
H$\beta$ full-width-at-half-maximum (FWHM) of $700\pm140$ km s$^{-1}$,
and the highest Eddington ratio of 12.9 among the 110 AGN in Grupe et
al. (2004) (also see Grupe et al. 2011). Paper-I showed that the X-ray
spectral-timing properties of this source are similar to \pg12, but
with even steeper 2-10~keV spectral index potentially indicating
higher $L/L_{\rm Edd}$ (e.g. Shemmer et al. 2006). Here we collect archival
data (Section 2) and show that it has an optical spectrum which is
very similar to \1h07, with extremely weak [O {\sc iii}] and strong Fe {\sc ii}
(Section 3). The full multi-wavelength spectral energy distribution
(SED) is presented in Section 4, where the extremely low Galactic
extinction of $N_{\rm H}=7.45\times10^{19}$ cm$^{-2}$ (Kalberla et
al. 2005) and low intrinsic extinction (Paper-I) mean that the spectrum is uniquely well defined.
The high-quality {\it HST} spectra gives UV data up to $\sim 10$~eV in the
rest frame of \rxj04, while the ROSAT data extends the X-ray spectrum
from \xmm\ down to 0.1~keV, so the bolometric luminosity is
constrained to better than 20\%.  Section 5 shows constraints on short
and long-term multi-wavelength variability of this SED.  Section 6
shows physical models of the accretion flow for the estimated black
hole mass and mass accretion rate. We clearly show that the mass
accretion rate through the outer disc is super-Eddington, and that
this does not convert to observed luminosity as expected for a
geometrically thin disc. Instead, at least half of the expected
luminosity is lost, probably powering a wind. We build a 
global picture of accretion flows in
all super-Eddington NLS1s and compare these NLS1s with weak
emission-line quasars at high redshifts. The final section summarises
the main results of this paper. Throughout this paper we adopt a flat
universe model for the luminosity distance with the Hubble constant
H$_{0} = 72$ km s$^{-1}$ Mpc$^{-1}$, $\Omega_{\Lambda} = 0.73$ and
$\Omega_{\rm M} = 0.27$.

\begin{table}
\centering
   \caption{Multi-wavelength data of \rxj04\ analysed in this work. `Exp' is the exposure time. The {\it WISE} observation (denoted by $\dagger$) was from 2010-01-28 to 2010-01-30, consisting of 37 separate exposures. The {\it WISE} observation on 2010-08-06 consisted of 2 separate exposures, but we cannot find the exposure time for {\it WISE} and {\it 2MASS}.}
     \begin{tabular}{@{}lccc@{}}
     \hline
    Instrument & Obs-Date & Exp & Waveband \\
     & & (ks) &\\
    \hline
    \xmm\ EPIC/OM & 2016-02-12 & 120 & X-ray/UV/Optical \\
    {\it ROSAT} PSPCB& 1997-02-20 & 1.3 & Soft X-ray \\
    {\it ROSAT} PSPCB& 1997-02-20 & 0.7 & Soft X-ray \\
    {\it ROSAT} PSPCB& 1997-02-26 & 0.6 & Soft X-ray \\
    {\it Swift} XRT/UVOT & 2006-01-06 & 5.4 & X-ray/UV/Optical \\
    {\it Swift} XRT/UVOT & 2006-04-13 & 4.0 & X-ray/UV/Optical \\
    {\it Swift} XRT/UVOT & 2006-04-15 & 4.3 & X-ray/UV/Optical \\
    {\it Swift} XRT/UVOT & 2006-05-18 & 3.4 & X-ray/UV/Optical \\
    {\it Swift} XRT/UVOT & 2012-05-09 & 1.0 & X-ray/UV/Optical \\
    {\it Swift} XRT/UVOT & 2012-05-14 & 1.5 & X-ray/UV/Optical \\
    {\it Swift} XRT/UVOT & 2014-04-05 & 1.0 & X-ray/UV/Optical \\
    {\it Swift} XRT/UVOT & 2014-04-17 & 0.9 & X-ray/UV/Optical \\
    {\it HST} COS & 2010-02-07 & 2.2 & UV (G130M) \\
    {\it HST} COS & 2010-02-07 & 1.0 & UV (G130M) \\
    {\it HST} COS & 2010-02-07 & 2.0 & UV (G130M) \\
    {\it HST} COS & 2010-02-07 & 1.0 & UV (G130M) \\
    {\it HST} COS & 2010-02-07 & 2.0 & UV (G130M) \\
    {\it HST} COS & 2010-02-07 & 0.8 & UV (G160M) \\
    {\it HST} COS & 2010-02-07 & 3.7 & UV (G160M) \\
    {\it HST} COS & 2010-02-07 & 2.2 & UV (G160M) \\
    {\it HST} COS & 2010-02-07 & 2.2 & UV (G160M) \\
    {\it HST} COS & 2010-05-26 & 4.2 & UV (G285M) \\
    {\it ESO 1.52 m Telescope} & 1999-09-14 & 2.7 & Optical \\
    {\it WISE} & 2010-01-30$^{\dagger}$ & -- & Infrared (Band 1-4) \\
    {\it WISE} & 2010-08-06 & -- & Infrared (Band 1-4) \\
    {\it 2MASS} & 1999-11-05 & -- & Infrared (J, H, K)\\
    \hline
     \end{tabular}
\label{tab-obs}
\end{table}

\section{Multi-wavelength Observations}
\label{sec-obs}
\subsection{Data Collection}
The latest 133 ks \xmm\ (Jansen et al. 2001) data of \rxj04\ was
obtained on 2016-02-12 (PI: C. Jin).  During the observation, all the
three European Photon Imaging Cameras (EPIC) (pn, MOS1, MOS2) were
operating in the {\it Imaging} mode. The Reflection Grating
Spectrometer (RGS) cameras were in the {\it Spectroscopy} mode. The
Optical Monitor (OM) was in the {\it Imaging+Fast} mode with about 10
ks exposure in each of the V, B, U, UVM2, UVW2 bands and about 70 ks
continuous exposure in the UVW1 band.

We also searched various public archives for existing multi-wavelength observations of \rxj04, and found a collection of datasets from infrared (IR) to hard X-rays (Table~\ref{tab-obs}):\\
(1) In the X-ray band \rxj04\ was observed by \rosat\ PSPCB in 1997 with three short exposures, which extend the X-ray spectral coverage down to $\sim$0.1 keV. \rxj04\ was also observed by {\it Swift} in 2006, 2012 and 2014 with totally nine observations, including five observations in 2006, two observations in 2012 and two observations in 2014. One observation in 2006 is excluded because its exposure time was very short (only 429 s). The remaining eight observations were conducted with the X-ray telescope (XRT) in the photon counting (PC) mode and simultaneous exposures in the UV/Optical Telescope (UVOT). The observation on 2014-4-17 has only U-band exposures, while all the other observations have exposures in all six UVOT bands (i.e. U, B, V, UVW1, UVM2, UVW2). Since all the {\it Swift} exposures are much shorter than the \xmm\ observation which provides high signal-to-noise (S/N) spectra in the same X-ray band, we only use the {\it Swift} data to study long-term variability. \\
(2) In the optical band we use the highest quality optical spectrum of \rxj04, which was obtained by the European Southern Observatory (ESO) 1.52 m telescope in 1999 (Grupe et al. 2004a).\\
(3) In the UV band, Hubble Space Telescope ({\it HST}) observed
\rxj04\ in 2010 with the Cosmic Origins Spectrograph (COS) instrument
(PI: J. Green) in order to obtain high S/N detections of the
emission/absorption-line system in the near/far-UV band. There are
also {\it WISE} and {\it 2MASS} photometric points in the near
infrared (IR) band. Note that these {\it WISE} observations were
obtained only one weak earlier than the {\it HST} observations on
2017-02-07, so they can be considered as nearly simultaneous
observations in the IR and UV bands. Since \rxj04\ has very low
extinction along the line of sight, these datasets provide direct
multi-wavelength information for the super-Eddington accretion flow in
\rxj04, which we present in the following sections.

\subsection{Data Reduction}
For the \xmm\ observation, we reduced the data using the {\sc SAS}
software (v15.0.0) and the latest calibration files. High background
periods were subtracted and no photon pile-up was found (see Paper-I
for a complete description). In this paper we only use the EPIC-pn
data with good events (FLAG=0) and PATTERN $\le$ 4. The source
spectrum and light curve were extracted from a circular region of 80
arcsec centred at \rxj04, with the background measured from a nearby
source-free region of the same size. OM light curves were created by
the {\tt omfchain} task, with the background measured from the image
data and subtracted automatically. For every OM filter the mean source
count rate was extracted from the OM source-list file and put into the
template file `om\_filter\_default.pi', which was then combined with
the latest canned response file using the {\tt grppha} task for later
spectral analysis\footnote{We notice that this OM data preparation
  procedure is now integrated into the {\sc SAS} {\tt om2pha} task,
  which produces the same spectral file.} (also see Jin et al. 2013, hereafter: J13).

We downloaded the {\it ROSAT} data from the High Energy Astrophysics Science Archive Research Center (HEASARC) data archive and followed the standard data reduction procedure\footnote{http://heasarc.gsfc.nasa.gov/docs/rosat/ros\_xselect\_guide/} and used {\sc Xselect} (v2.4d) to extract source and background spectra. The {\tt pcarf} task was run to generate auxiliary files and response matrices. Since the three {\it ROSAT} observations were all conducted within one week (see Table~\ref{tab-obs}), we combined their spectra using the {\tt addspec} task ({\sc FTOOL} v6.19) to maximise the S/N.
 
{\it Swift} data were also downloaded from the HEASARC data archive. Standard data reduction procedures were followed according to the official threads\footnote{http://www.swift.ac.uk/analysis/}, using the softwares provided in the {\sc HEASOFT} (v6.19) package and latest calibration files in the calibration database (CALDB). For the XRT data, the {\tt xrtpipeline} task ({\sc HEASOFT} v6.19) was used to reprocess the data and produce the event file. Then the images, source and background spectra were extracted within the {\sc Xselect} environment, including a standard check for the photon pile up (not found). Auxiliary files were produce using the {\tt xrtmkarf} task for spectral analysis. For the UVOT data, the {\tt uvotimsum} and {\tt uvot2pha} tasks were used to sum exposures and create spectral files for every filter. 

The {\it HST} spectra of \rxj04\ have been well calibrated and analysed by Danforth et al. (2016) in order to probe the foreground intergalactic medium by using the well resolved absorption lines (also see Keeney et al. 2013). The calibrated and combined spectra were downloaded directly from the Mikulski Archive for Space Telescopes (MAST), where a sophisticated line detection procedure has also been performed (Danforth et al. 2016).

\begin{figure}
\includegraphics[scale=0.42]{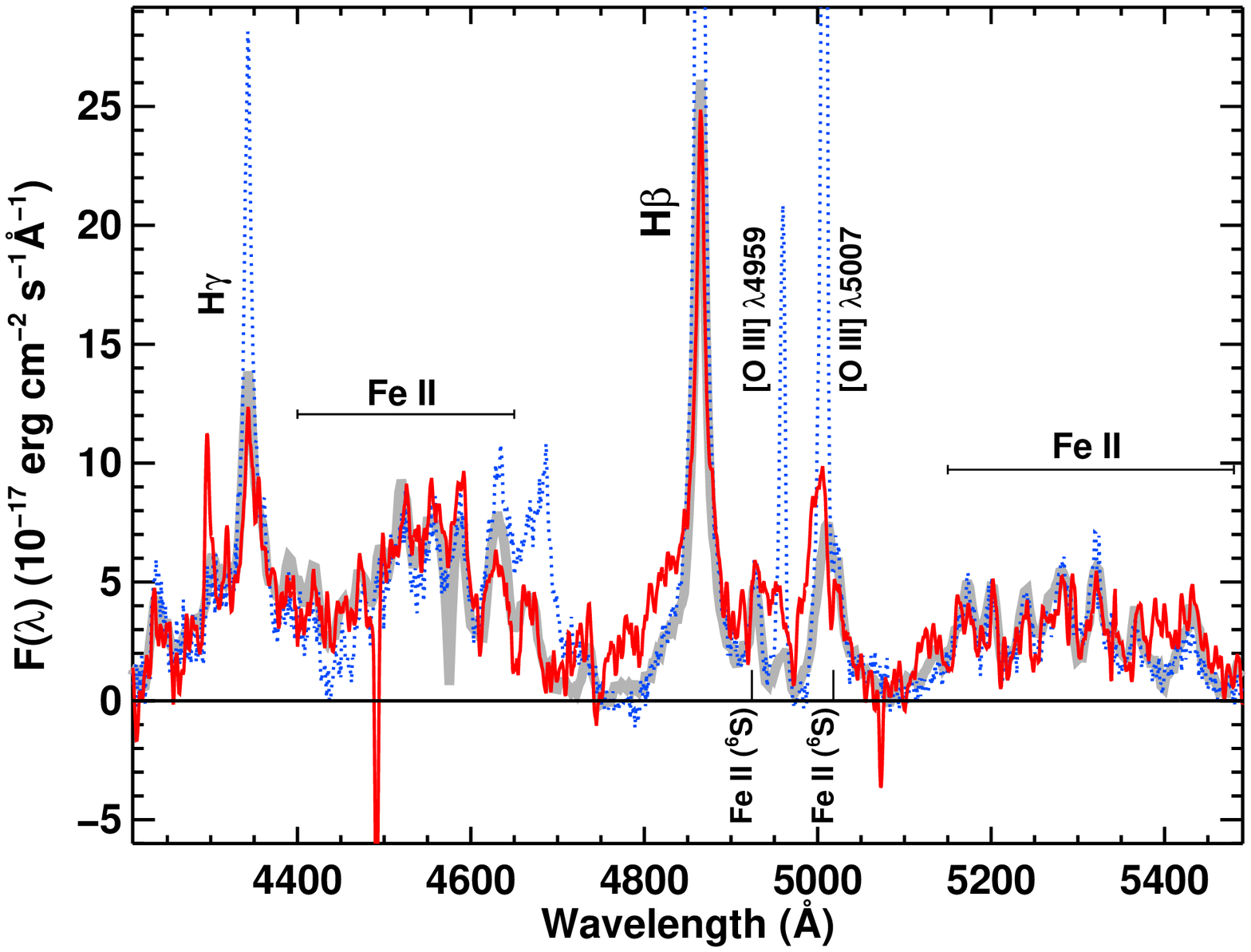}
\caption{Comparison of optical emission lines between \pg12\ (dotted blue), \1h07\ (thick gray) and \rxj04\ (solid red, smoothed by a factor of 5 to increase the visibility). Local underlying continua have been subtracted for every source. The line intensities of \pg12\  and \1h07\ are rescaled to match the Fe {\sc ii} lines of \rxj04\ within 5100-5500\AA. Both \rxj04\ and \1h07\ show much weaker NLR lines than \pg12.}
\label{fig-optcompare}
\end{figure}

\section{Optical/UV Spectral Constraints}
\label{sec-optspec}
\subsection{The Redshift of \rxj04}
An optical spectrum can provide direct constraints on the black hole
mass and mass accretion rate, and also provide information about
different emission line regions (NLR and broad line region: BLR). The
redshift of \rxj04\ given in the NASA/IPAC EXTRAGALACTIC DATABASE
(NED) is 0.243 with no quoted error, which was originally reported by
Thomas et al. (1998). Since \rxj04\ has very weak NLR lines and the
original spectrum has low resolution (FWHM=30\AA), this redshift may
not be very accurate. In order to obtain a more accurate redshift
estimate, we match the optical lines of \rxj04\ with those of \pg12,
whose redshift is precisely measured to be $z=0.0482\pm0.0008$ from
its strong NLR lines. In this way, we find $z=0.242$ to be a more
accurate redshift for \rxj04, which results in a good matching of most
optical lines with \pg12, except the [O {\sc iii}]$\lambda$4959, 5007
doublets which are clearly weaker, broader and more blueshifted  in
\rxj04\ than in \pg12\  (Fig.~\ref{fig-optcompare}). Next we add
\1h07\  for comparison ($z=0.0398$, DJ16). It is clear that \rxj04\ is
more similar to \1h07\  in terms of the strength of the [O {\sc iii}]
lines relative to Balmer lines, but these lines in \rxj04\ are also more blueshifted than those in \1h07. Note that if we adopt the NED redshift value of $z=0.243$, \rxj04\ would exhibit significant blueshifts in all NLR and BLR lines relative to \pg12, which is not likely to be correct.

\begin{figure}
\includegraphics[bb=-10 120 520 720, scale=0.38]{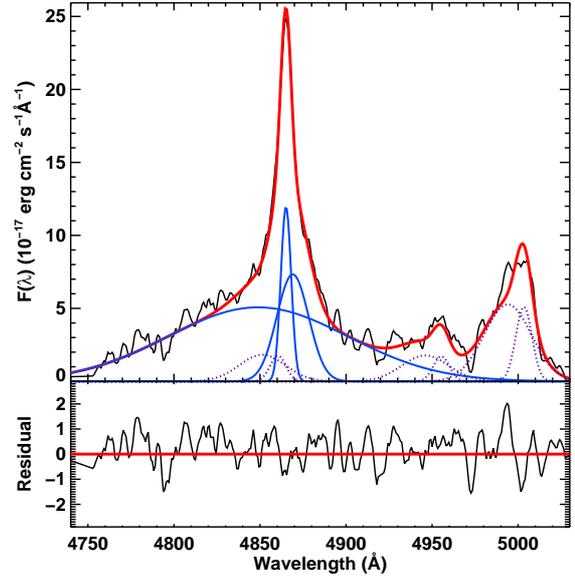}
\caption{Multi-Gaussian fit to the H$\beta$ line and [O {\sc iii}]$\lambda$4959/$\lambda$5007 doublets in \rxj04. The spectrum is smoothed by a factor of 5 to increase the visibility. The Fe {\sc ii} lines have been subtracted. Dotted profiles show the Gaussian components for the NLR lines. Solid blue profiles show the Gaussian components for the BLR lines. See Section~\ref{sec-optspec} for detailed fitting strategy. The lower panel shows the residuals (i.e. data - model).}
\label{fig-hbfit}
\end{figure}

\begin{table}
 \centering
   \caption{The best-fit parameters for the line profiles of [O {\sc iii}]$\lambda$5007 and H$\beta$ (Fig.~\ref{fig-hbfit}). Line-shift is relative to the vacuum wavelength, which is 5008.24\AA\ for [O {\sc iii}]$\lambda$5007 and 4862.68\AA\ for H$\beta$, with negative values indicating blueshifts. Errors are for 1 $\sigma$ confidence level. EW is the equivalent width. An extra systematic uncertainty of $\sim300$ km s$^{-1}$ due to the redshift uncertainty should also be considered in all the line-shift measurements. NC is the narrow component in H$\beta$ with the same line profile as [O {\sc iii}]$\lambda$5007.}
     \begin{tabular}{lccc}
\hline
  Component & Line-shift & FWHM & EW\\
  & (km~s$^{-1}$) & (km~s$^{-1}$) & (\AA) \\
  \hline
  \multicolumn{4}{c}{[O {\sc iii}]$\lambda$5007}\\
  Gaussian-1 & $-290\pm30$ &  $660\pm80$ &$2.0\pm0.4$ \\
  Gaussian-2 & $-860\pm60$ & $1940\pm70$ & $6.0\pm0.4$\\
  \multicolumn{1}{c}{total} & -- & $1360\pm180$ &$8.0\pm0.6$\\
  \hline
  \multicolumn{4}{c}{H$\beta$}\\
  Gaussian-1&$150\pm20$   &  $440\pm60$ &$2.8\pm0.6$\\
  Gaussian-2&$390\pm110$ &  $1340\pm140$ &$5.2\pm1.0$\\
  Gaussian-3&$-860\pm70$   &  $7580\pm240$ &$20.4\pm0.5$\\
  \multicolumn{1}{c}{NC} & -- & -- & $0.6\pm0.3$ \\
  \multicolumn{1}{c}{total} & -- & $850\pm170$ &$29.0\pm1.3$\\
\hline
   \end{tabular}
 \label{tab-hbfit}
\end{table}

\subsection{The NLR and BLR Outflows}
AGN optical spectra often show abundant emission lines from both NLR and BLR, whose line profiles can be used to infer the physical properties and dynamics of these distinct emission line regions (e.g. Blandford \& McKee 1982; Antonucci 1993; Peterson et al. 2014; Dopita et al. 2015). We performed a multi-gaussian line profile fitting to quantify the parameters of [O {\sc iii}] and H$\beta$ lines. Firstly, the Fe {\sc ii} lines were fitted between 4000-5500\AA\ after subtracting a local continuum and masking out all other strong lines in between. The Fe {\sc ii} template used in the fitting consists of the Fe {\sc ii} blends in I Zw 1 and four extra Fe {\sc ii} line groups (P, F, S, G), with the line-width and relative intensities determined by the fitting (Kova\v{c}evi\'{c}, Popovi{\'c} \& Dimitrijevi{\'c} 2010; Shapovalova et al. 2012). After subtracting the best-fit Fe {\sc ii} template, we used multiple Gaussian components to fit the [O {\sc iii}]$\lambda$4959/5007 doublets and H$\beta$ line, simultaneously. The [O {\sc iii}]$\lambda$5007 line profile requires two gaussian components. [O {\sc iii}]$\lambda$4959 was assumed to have an identical profile as [O {\sc iii}]$\lambda$5007 with a fixed atomic flux ratio of 1:3. We also assumed that the narrow component of H$\beta$ line has the same profile as [O {\sc iii}]$\lambda$5007. Besides the narrow component, we found it necessary for H$\beta$ to have three extra Gaussian components: one fitting the narrow peak and two fitting the broad base. The fitting was performed using the MPFITEXPR program (Markwardt 2009) in the Interactive Data Language ({\sc IDL} v8.4), and the results are shown in Fig.~\ref{fig-hbfit} and Table~\ref{tab-hbfit}. The total FWHM of H$\beta$ was measured directly from the combined profile of best-fit Gaussian components excluding the narrow component, with its error determined by the Monte Carlo method. We point out that the redshift uncertainty of $\sim0.001$ corresponds to a velocity of 300 km s$^{-1}$, which certainly affects the measurement of line-shift in every emission line.

From the line fitting results in Table~\ref{tab-hbfit}, we find each of the [O {\sc iii}] double lines is dominated by a broad Gaussian component with FWHM $=1940\pm70$ km s$^{-1}$, blueshifted by $860\pm60$ km s$^{-1}$ and comprises 75\% of the total line flux, making the whole line profile very broad. This suggests that most of the NLR clouds emitting [O {\sc iii}] are outflowing. The fitting requires very little NLR component of the same profile as [O {\sc iii}] in the H$\beta$ line, but an extra component is required to fit the narrow peak of H$\beta$. This suggests that the H$\beta$ NLR has different properties from the [O {\sc iii}] NLR. Then the broad base of H$\beta$ is dominated by a broad (FWHM $=7580\pm240$ km s$^{-1}$) and blueshifted ($v=-860\pm70$ km s$^{-1}$) component. The total H$\beta$ FWHM is only $850\pm170$ km s$^{-1}$, which is consistent with previous measurements (Grupe et al. 2004a,b). Combining the results in Fig.~\ref{fig-optcompare} and \ref{fig-hbfit}, we can see that the BLR clouds in \rxj04\ show an outflow that is much stronger than in either \pg12\  or \1h07, suggesting a more powerful radiation field in \rxj04. Compared to the strong blueshift in the NLR lines, only part of the H$\beta$ line profile is blueshifted, suggesting that the BLR clouds are likely optically thick, and so the radiation acceleration only causes an outflow in the lower density surface of the BLR clouds (Baskin, Laor \& Stern 2014a).

\begin{figure}
\centering
\begin{tabular}{c}
\includegraphics[bb=25 144 878 810, scale=0.275]{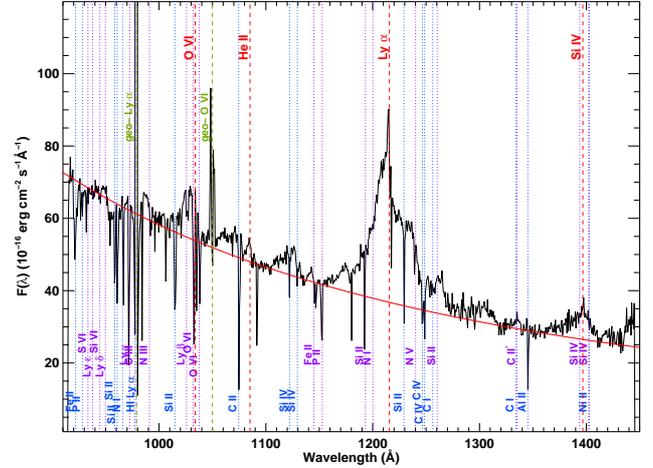}
\end{tabular}
\caption{The combined {\it HST} COS spectrum (G130M + G160M) of \rxj04, rebinned with 0.6 \AA\ per bin to increase the S/N. The spectrum has been de-reddened for a Galactic reddening of $E(B-V)=0.013$ and de-redshifted for z=0.242 to the AGN rest-frame. The red solid line is the best-fit power law to the continuum with $\alpha=2.35\pm0.02$. Most of the absorption lines can be identified as Galactic absorption lines (blue) or host galaxy absorption lines (magenta). We also find two geo-coronal lines (dark green) and four emission lines (red) intrinsic to \rxj04.}
\label{fig-uvspec}
\end{figure}

\subsection{Mapping the Accretion Flow in UV}
\label{sec-uvspec}
For AGN with high mass accretion rates, their UV emission is often dominated by the emission from the outer accretion disc (hundreds of $R_{\rm g}$), and so the shape of the UV underlying continuum can be used to map the outer accretion flow. Fig.~\ref{fig-uvspec} shows the combined spectrum of \rxj04\ from all previous COS exposures (G130M \& G160M) (Danforth et al. 2016). The spectrum extends down to 1120 \AA\ in the observer's frame, or 900 \AA\ in the rest-frame of \rxj04, which corresponds to a radius of 190 $R_{\rm g}$ in a standard accretion disc model (Shakura \& Sunyaev 1973, hereafter: SS73) with $M=1\times10^{7}~\rm M_{\odot}$ and $\dot{m}_{\rm out}~{\equiv}~\dot{M}_{\rm out}/M~=~5.9$ (see Section~\ref{sec-bhmass}). There are many narrow absorption lines from the Milky Way and the host galaxy, as well as broad emission lines (Ly$\alpha$, Si {\sc iv}, He {\sc ii}, O {\sc vi}) due to the AGN activity in \rxj04. No broad absorption lines can be identified, which indicates a relatively small inclination angle and a clear line-of-sight to the accretion flow.

We used a power law ($f_{\lambda}\propto\lambda^{-\alpha}$) to fit the underlying continuum after carefully masking out all strong emission and absorption lines, and found $\alpha=2.25\pm0.02$ before de-reddening the spectrum. After applying the correction for the Galactic reddening of $E(B-V)=0.013$ (assuming $E(B-V)=1.7\times10^{-22}N_{\rm H}$, Bessell 1991) using the reddening curve of Milky Way in Pei (1992), the best-fit power law slope increases to $\alpha=2.35\pm0.02$ (Fig.~\ref{fig-uvspec}, red solid line), which is fully consistent with the spectral slope of a standard thin disc model. Therefore, the unobscured UV continuum from {\it HST} COS clearly indicates the accretion flow in \rxj04\ behaves like a standard accretion disc down to 190 $R_{\rm g}$ for the above mass and mass accretion rate. Previous studies of AGN composite UV spectra show a flatter spectral shape than a standard accretion disc spectrum at $\lambda>1000$\AA, as well as an abrupt turnover at $\sim$1000\AA\ (e.g. Zheng et al. 1997; Telfer et al. 2002; Shang et al. 2005; Barger \& Cowie 2010; Shull, Stevans \& Danforth 2012). We do not find any spectral shape change at 1000\AA\ in \rxj04\ (Fig.~\ref{fig-uvspec}), but we cannot rule out the existence of such a turnover below 900 \AA.

\begin{figure*}
\centering
\begin{tabular}{c}
\includegraphics[bb=54 216 702 750, scale=0.56]{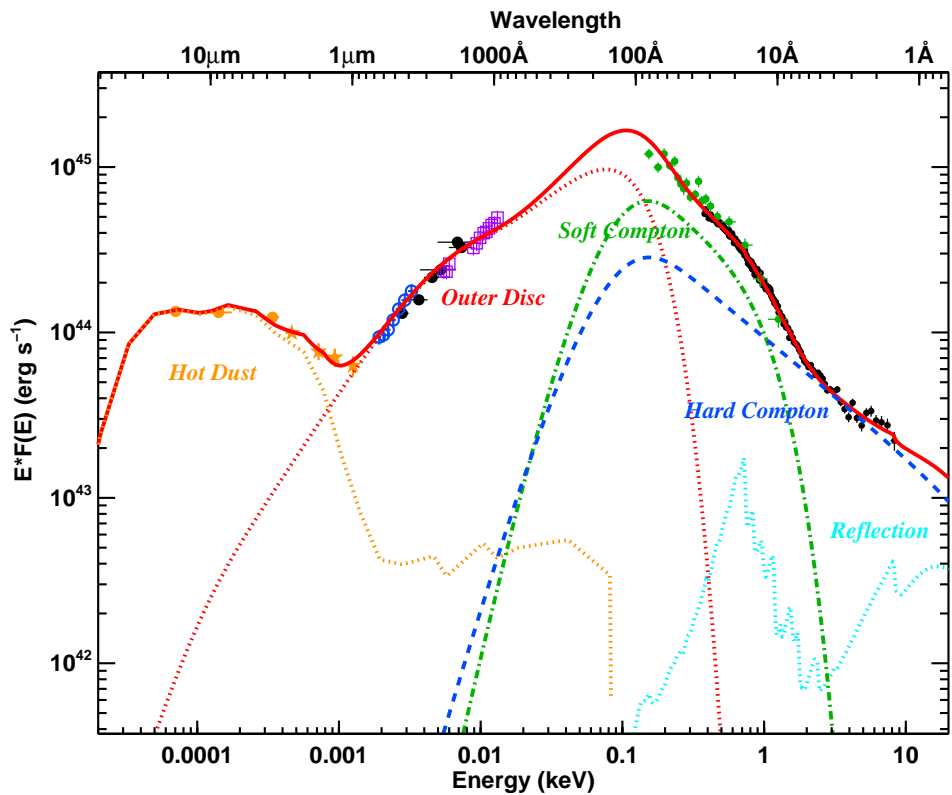} \\
\end{tabular}
\caption{The broadband SED of \rxj04, assuming an inclination angle of $30\degr$. The data consist of \xmm\ EPIC-pn spectrum and OM photometric points (black), a combined {\it ROSAT} spectrum (green points in the X-ray, scaled up by 3\%), continual points from the {\it HST} COS spectra (magenta in the optical/UV, scaled down by 23\%), the optical spectrum from Grupe et al. (2004a) (blue, scaled down by 20\%), and the IR photometric points including WISE Band 1-4 (orange circles, scaled down by 23\%) and 2MASS J, H, K (orange stars, scaled down by 23\%). Red solid curve is the best-fit SED model, comprising an accretion disc component (red dotted curve), a soft X-ray Comptonisation component (green dash-dot curve), a hard X-ray Comptonisation component (blue dash curve), a weak reflection component (cyan dotted curve) and a hot dust component (orange dotted curve). Note that this broadband SED model does not consider any energy loss due to the disc wind or advection.}
\label{fig-sed}
\end{figure*}

\section{Broadband SED Analysis}
\label{sec-sed}
A broadband SED can be used to constrain the accretion flow and measure the key parameters such as the bolometric luminosity and Eddington ratio (e.g. Jin et al. 2012a, Done et al. 2013; DJ16). In order to construct the broadband SED of \rxj04, we make use of the combined \rosat\ spectrum in the soft X-ray hand, \xmm\ EPIC-pn spectrum in the 0.3-10 keV band, \xmm\ OM data in the UV/optical, continua measured from the combined {\it HST} spectrum and the single-epoch optical spectrum, as well as {\it WISE} and {\it 2MASS} photometric points in the IR band. We imported all these data into {\sc Xspec} (v12.9.0o, Arnaud 1996) and performed the broadband SED fitting. The {\sc optxconv} model is used to model the broadband SED from optical to hard X-rays, which comprises an accretion disc component, a soft X-ray Comptonisation component and a hard X-ray Comptonisation component. The viewing angle effect and relativistic effects are also included in it (Done et al. 2013). Our detailed X-ray study in Paper-I shows that there could also be a weak reflection component, so we include a {\sc kdblur}$\times${\sc rfxconv}$\times${\sc nthcomp} component in the model (see Paper-I for detailed explanations). The Galactic and intrinsic extinctions are both modelled with the {\sc tbnew} model using cross-sections of Verner et al. (1996) and interstellar medium (ISM) abundances of Wilms, Allen \& McCray (2000). Galactic and intrinsic reddening are modelled with the {\sc (z)redden} model, assuming $E(B-V)=1.7\times10^{-22}N_{\rm H}$ (Bessell 1991). The IR data are fitted with a hot dust template from Silva, Maiolino \& Granato (2004). A Sa-type host galaxy template from the SWIRE library (Polletta et al. 2007) is also included to model the host galaxy starlight in the optical/UV band. We adopt z=0.242, an inclination angle of $30\degr$ and a zero spin for the SED fitting.

The normalisation discrepancy between non-simultaneous observations from different instruments is a major issue when fitting a broadband SED. This discrepancy consists of both aperture effect and variability of the source. \rxj04\ is a point-like source from IR to X-rays, so the aperture effect between different instruments is minimised. Regarding the variability issue, we have simultaneous optical/UV and X-ray data from \xmm. The {\it WISE} IR and {\it HST} UV observations are also nearly simultaneous (Table~\ref{tab-obs}). \rxj04\ is also a high mass accretion rate QSO, which implies that its IR/optical/UV emission should be relatively stable (e.g. Ai et al. 2013; Meusinger \& Weiss 2013). Therefore, the influence of  source variability should also be small. We multiply a free scaling factor to the entire SED model to account for small normalisation discrepancies.

Our SED model produces a good fit to all the multi-wavelength data with $\chi^2_v=1089/712$ (Fig.~\ref{fig-sed}). Small normalisation discrepancies are found relative to the \xmm\ data, which is a factor of 1.31 for the {\it HST} spectra and IR data, 1.24 for the optical spectrum, and 0.97 for the {\it ROSAT} spectrum. The discrepancy in the optical/UV band would be smaller if we were able to correct for the contribution of emission line fluxes in the OM photometric points. However, because there was no simultaneous optical/UV spectra for the OM data, we have not attempted to do so. The soft and hard X-ray data are well fitted by the two Comptonisation components plus the weak reflection component, with the accretion disc component extending slightly into the soft X-ray band. No intrinsic extinction is required by the fitting ($N_{\rm H}~<~10^{19}$cm$^{-2}$). These results are all consistent with our X-ray study in Paper-I. The optical/UV data are well fitted by the accretion disc spectrum without requiring any intrinsic de-reddening or a contribution from host galaxy star-light. The IR data are well fitted by the hot dust template alone. The fitting finds $M=1.8\times10^{7}~\rm M_{\odot}$ and $L/L_{\rm Edd}=1.8$ for a zero spin black hole. The corona radius $R_{\rm cor}$ is found to be 29 $R_{\rm g}$, within which all the disc energy is required to be dissipated into the two Comptonisation components. Although a better fit with $\Delta\chi^2=7$ can be found for $R_{\rm cor}=100~R_{\rm g}$, the disc component can no longer extend into the soft X-ray band for such a large $R_{\rm cor}$, which is not consistent with our X-ray study discussed in Paper-I, therefore we rejected this fit. The hard X-ray Comptonisation contains 31\% of the total Comptonisation energy and has a steep slope of $\Gamma=2.71$. The soft X-ray Comptonisation has an electron energy of 0.25 keV and optical depth of 14.9, broadly consistent with the results in Paper-I.

The {\sc optxconv} model is self-consistent in terms of energy budget, although in the total SED model it does not include the energy in reflection which is very little. However, a major issue is that it does not include the energy lost due to a disc wind and/or advection, which mainly affects super-Eddington sources like \rxj04\ (Jin, Done \& Ward 2016; DJ16). This means that {\sc optxconv} can over-estimate the disc luminosity at small radii, thereby producing a UV bump that is too strong when compared with the observations. Then the fitting will require an increase in the black hole mass in order to reduce the disc temperature and the strength of the UV bump, and so the best-fit black hole mass is then over-estimated. However, the assumption of zero spin can lead to an under-estimate of the black hole mass. For example, if we change the spin parameter to its maximal value of 0.998, the radiative efficiency increases to 0.321 (DJ16), and the UV bump becomes stronger and shifts to higher energy. Then the best-fit black hole mass increases to $3.5\times10^{7}~\rm M_{\odot}$ (i.e. up by a factor of 2 compared with that for zero spin) with an Eddington ratio of 2.2 in order to maintain a good fit. Although the true spin of \rxj04\ cannot be determined from the existing data, previous studies show that SMBH with $\sim10^{7}~\rm M_{\odot}$ may have a chaotic/episodic accretion history which leads to a low spin (e.g. King, Pringle \& Hofmann 2008; Miniutti et al. 2009; Fanidakis et al. 2011; but see Orban de Xivry et al. 2011). So any bias caused by the assumption of zero spin may be smaller than that resulting from neglecting the presence of the disc wind and advection. Finally, we emphasise that the SED parameters reported above should be considered only as indicative for the intrinsic properties of the source, particularly noting that the mass is likely to be over-estimated (also see Section~\ref{sec-bhmass}).

\begin{figure}
\includegraphics[bb=70 360 558 864, scale=0.50]{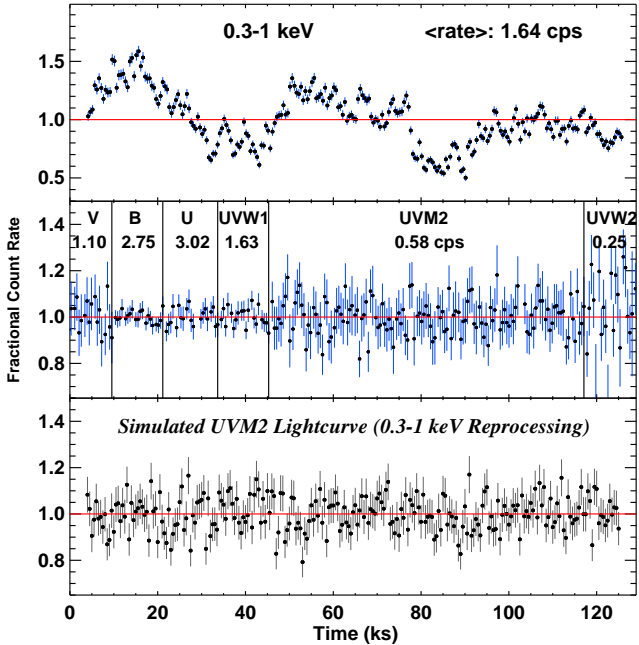}
\caption{Short-term variability of \rxj04\ revealed by the 500 s binned, background subtracted light curves from EPIC-pn in the 0.3-1 keV band and the simultaneous exposures in the six OM bands. The labelled number indicates the mean source count rate in the unit of counts per second (cps). The bottom panel shows our simulation resulting from the reprocessing of the 0.3-1 keV light curve emerging in the UVM2 band, whose intrinsic RMS is only 0.22\%.}
\label{fig-lc}
\end{figure}

\section{Multi-Wavelength Variability}
\subsection{Short-term Variability}
By virtue of the unique capability of \xmm, which allows the long and simultaneous observation in the X-ray and UV/optical bands with high time-resolution, the multi-wavelength short-term variability of \rxj04\ can be revealed clearly (Fig.~\ref{fig-lc}). In Paper-I we have performed a full analysis of the short-term variability in the X-ray band and reported the Root-Mean-Square (RMS) fractional variability of $22.4\pm0.4$\%, $28.7\pm0.4$\% and $47.5\pm1.1$\% in the 0.3-1, 1-2 and 2-10 keV bands, respectively. Compared to the strong X-ray variability, Fig.~\ref{fig-lc} shows that the optical/UV emission from \rxj04\ contains very little short-term variability. The intrinsic RMS variability is found to be $10.7\pm5.8$\% in UVW2, and is consistent with 0 in the other OM bands. We also searched for the covariance between UVM2 and X-rays light curves by calculating the cross-correlation function (CCF), but the variability in UVM2 is too small to allow any significant detection of UV/X-ray coherence or time-lag.

The SED in Fig.~\ref{fig-sed} shows that the flux contained in the soft X-rays is comparable to the flux in the optical/UV band, so if some of the highly variable soft X-rays illuminate the outer disc and are reprocessed as part of the disc emission, it may cause the optical/UV light curve to vary as well. This X-ray reprocessing mechanism has been proposed for many AGN in order to explain the commonly observed optical/UV reverberation lag in long-term monitoring campaigns of sub-Eddington AGN (e.g. Mason et al. 2002; Ar\'{e}valo et al. 2005; Alston, Vaughan \& Uttley 2013; Lohfink et al. 2014; Edelson et al. 2015; Buisson et al. 2017; but see Gardner \& Done 2017). However, the short-term variability is more easily smeared in the outer disc, leading to a much weaker reverberation signal which is difficult to detect (Smith \& Vaughan 2007; Robertson et al. 2015).

In order to understand how much variability is expected in UVM2 if the
soft X-rays within 0.3-1 keV is indeed reprocessed in the outer disc,
we perform a simulation of this variability transmission using
methods similar to those employed in Gardner \& Done (2017). For the
simulation we adopted a $10^{7}~\rm M_{\odot}$ black hole with a mass
accretion rate of 5.9 (see Section~\ref{sec-mdot}). The soft X-ray
source is located at 30 $R_{\rm g}$ above the black hole on the spin
axis (similar to the `lamp-post' model geometry). The soft X-rays
illuminate the outer disc, increasing the temperature and luminosity
of every annulus. Then the emission from each annulus contains a
constant disc component and a variable reprocessed component. The
observed flux in UVM2 (bandpass: $2310\pm240$ \AA, \xmm\ Users
Handbook) contains the emission from a range of annuli in the outer
disc spanning hundreds of $R_{\rm g}$. We use the PSD of 0.3-1 keV and
the algorithm of Timmer \& Konig (1995) to simulate more light curve
segments, which are used as input for the simulation.
The bottom panel of Fig.~\ref{fig-lc} shows the resultant UVM2 light
curve with the same S/N as the real observation. 

The suppression of reprocessed soft X-ray variability are mainly due to three causes. These include the dilution by the intrinsic UV disc emission, which can be considered as a stable component compared to the reprocessed component; the transfer function at each radius, which takes into account different light travel time to different azimuths at each radius; the wide outer disc region spanning hundreds of $R_{\rm g}$ that contributes significant flux in the UVM2 band with different light travel time (see Gardner \& Done 2017). As a consequence, the simulated UVM2 light curve contains only 0.22\% intrinsic RMS variability, which is hardly detectable with the given S/N. Adding the hard X-ray emission to the reprocessing cannot increase the UV variability significantly, because its luminosity is one order of magnitude less than that of the soft X-ray emission, and the reprocessed hard X-ray variability suffers similar suppression.

We also note that the above result does not depend on the input black hole mass or mass accretion rate. This is because a standard thin disc model has $T^4\propto M\dot{M}/R^3$ (SS73) and monochromatic luminosity below the disc peak of $L_{\rm opt}\propto (M\dot{M})^{2/3}\propto (M^2\dot{m})^{2/3}$ (Laor \& Davis 2011, see also Section~\ref{sec-mdot}). Hence $T\propto L_{\rm opt}^{3/8} R^{-3/4}$ so that the radius in the disc which produces a given temperature is completely specified by the observed optical luminosity and does not depend on the model mass and mass accretion rate, hence the radius (e.g. $\sim$670 $R_{\rm g}$ for M = $10^{7}$ M$_{\odot}$ and $\dot{m}=5.9$) which produces a temperature which peaks in the UVM2 band is completely defined by the data.

\begin{figure}
\includegraphics[scale=0.5]{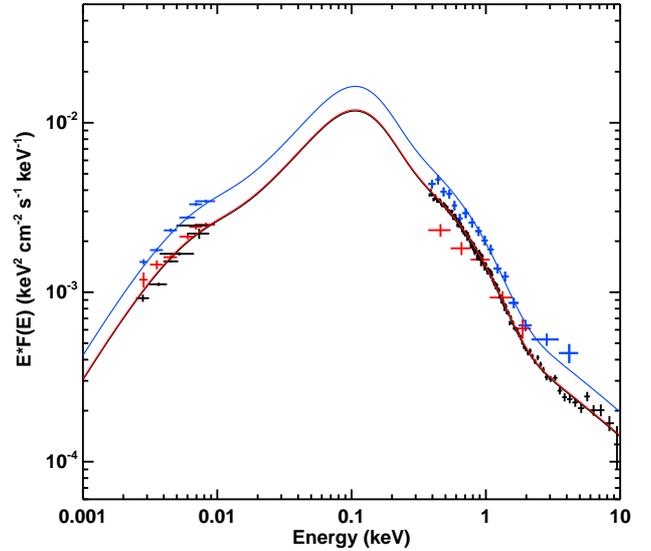}
\caption{The long-term multi-wavelength variability of \rxj04\ revealed by {\it Swift} XRT/UVOT (blue: combined from four observations in 2006; red: combined from four observations in 2012 and 2014) and \xmm\ EPIC-pn/OM (black: observed in 2016). The solid curves are the best-fit {\sc optxconv} models, with the blue model being 38\% higher than the red model and 40\% higher than the black model.}
\label{fig-sed-swift}
\end{figure}

\subsection{Long-term Variability}
\label{sec-longvar}
Although the short-term reprocessed X-ray variability cannot be detected in the outer disc emission, the fluctuation of mass accretion rate in the outer disc with much longer timescale can introduce a long-term variability, which may cause the accretion disc and corona emission to vary in a correlated pattern. \rxj04\ was monitored by the Catalina Survey from 2005-12-09 to 2012-12-18, and exhibited a factor of $\sim$2 variation in the V-band light curve, confirming the existence of a long-term variability in the outer disc.

The eight {\it Swift} observations of \rxj04\ between 2006 and 2014 (see Table~\ref{tab-obs}), combined with our new \xmm\ observation, provide the opportunity to test the long-term covariance between optical/UV and X-ray emission. However, all four observations obtained in 2006 show consistent fluxes, while the four observations in 2012 and 2014 also show similar flux but with low S/N due to the short exposure. Therefore, we combined the four observations in 2006 and the four observations in 2012-2014 to increase the S/N. The {\tt addspec} and {\tt fappend} tasks from {\sc FTOOL} (v6.19) were used to combine the XRT spectra and UVOT exposures, separately. We took the best-fit {\sc optxconv} model in Fig.~\ref{fig-sed} and multiplied it by a free constant parameter to fit the two combined {\it Swift} SEDs in 2006 and 2012-2014. Fig.~\ref{fig-sed-swift} shows that the best-fit SED model for the \xmm\ data can also fit the {\it Swift} data simply by adjusting the normalisation. The SEDs observed by {\it Swift} in 2006 and 2012-2014 are a factor of $1.40\pm0.02$ and $1.02\pm0.02$ as luminous as observed by \xmm\ in 2016, respectively. These results suggest a clear long-term covariance between the optical/UV disc emission and the X-ray corona emission, which is likely caused by the fluctuation of mass accretion rate in the outer disc. This fluctuation can change the luminosity of the accretion flow, but is probably not large enough to cause a significant change in the structure of the flow, and so it mainly affects the normalisation of the broadband SED rather than the overall shape.

\begin{figure}
\includegraphics[scale=0.42]{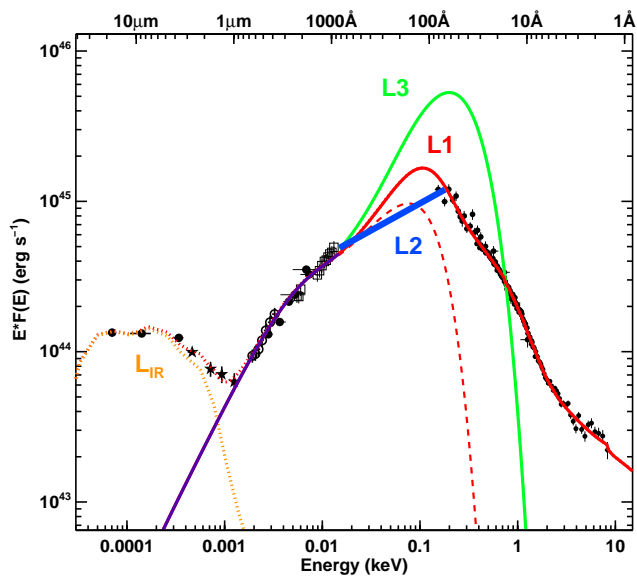}
\caption{The same broadband SED as in Fig.~\ref{fig-sed}. The red solid curve is the {\sc optxconv} plus reflection model with $M=1.8\times10^{7}~\rm M_{\odot}$ and $L_{1}/L_{\rm Edd}=1.8$ (the disc component is indicated by the red dash curve). An accretion disc spectrum of $M=7\times10^{6}~\rm M_{\odot}$ and $\dot{m}=12.1$ is added to match the optical/UV flux (green solid curve, with a bolometric luminosity of $L_{\rm 3}$), which clearly overshoots the soft X-ray data by a factor of up to 6.3. The thick blue line is a direct link from the UV data to the soft X-ray data, which provides a lower limit of the bolometric luminosity ($L_{\rm 2}$). $L_{\rm IR}$ is the luminosity in the hot dust component (orange dotted curve). We find $L_1 = 1.2 L_2 = 0.4 L_3  =10.8 L_{\rm IR}$ (see Section~\ref{sec-mdot}).}
\label{fig-sed2}
\end{figure}

\section{Discussion}
\subsection{Black Hole Mass Estimates}
\label{sec-bhmass}
The black hole mass of a SMBH is often measured from single-epoch optical spectra (e.g. Vestergaard 2002; McLure \& Jarvis 2002). For \rxj04\ Grupe et al. (2004a,b) measured the FWHM of H$\beta$ line to be 700$\pm140~km~s^{-1}$ and derived a black hole mass of $3.9\times10^{6} ~\rm M_{\odot}$ using the scaling relation reported by Kaspi et al. (2000). We revisit this black hole mass estimate using the results of our line fitting in Section~\ref{sec-optspec} and applying more up-to-date scaling relations in the literatures.

Our H$\beta$ line profile fitting gives a FWHM of $850\pm170$ km s$^{-1}$ (Table~\ref{tab-hbfit}), which is broadly consistent with previous work. We notice that the partial blueshift in the H$\beta$ line's base does not affect the FWHM of the line, because the FWHM is mainly determined by the two narrower components (see Fig.~\ref{fig-hbfit}). The flux measured at the rest-frame 5100{\AA} is $2.96\times10^{-16}$ erg cm$^{-2}$ s$^{-1}$ \AA$^{-1}$. With a luminosity distance of 1187.1 Mpc (Wright 2006) and assuming an isotropic source, we find the monochromatic luminosity at 5100\AA\ to be $L_{5100\AA}=2.55\times10^{44}$ erg s$^{-1}$. Then the black hole mass is estimated to be $9.4\times10^{6}~\rm M_{\odot}$ using the Vestergaard \& Peterson (2006) (VP06) relation or $6.7\times10^{6}~\rm M_{\odot}$ using the Woo \& Urry (2002) (WU02) relation. These mass estimates typically have a 1$\sigma$ systematic uncertainty of $\sim0.5$ dex. If we use the Marconi et al. (2010) (M08) scaling relation to correct for the radiation pressure within the BLR, the black hole mass would increase to the value of $1.0\times10^{8}~\rm M_{\odot}$. If we assume an inclination angle of $30\degr$ rather than isotropic, $L_{5100\AA}$ reduces to $1.47\times10^{44}$ erg s$^{-1}$, then the black hole mass reduces to $7.1\times10^{6}~\rm M_{\odot}$ for the VP06 relation, $4.6\times10^{7}~\rm M_{\odot}$ for the WU02 relation, and $5.9\times10^{7} ~\rm M_{\odot}$ for the M08 relation. But it has been reported that the M08 relation is likely to over-estimate the black hole mass, because the weakly ionised BLR clouds are like to be optically thick as suggested by the H$\beta$ line profile (see Section~\ref{sec-optspec}), and so the radiation pressure can only affect the surface of the BLR rather than dominating the entire region (Baskin, Laor \& Stern 2014b).

Another independent black hole mass estimate can be obtained using the correlation between the black hole mass and the hard X-ray variability, regardless of the details about the variability mechanism (Miniutti et al. 2009; Zhou et al. 2010; Ponti et al. 2012; Ludlam et al. 2015; Jin, Done \& Ward 2016). The 2-10 keV excess variance ($\sigma^{2}_{\rm rms}$) of \rxj04\ is $0.078\pm0.007$ for the 250s-binned 40 ks light curve segments from the \xmm\ observation; while for a 80 ks light curve segment, we find $\sigma^{2}_{\rm rms}=0.086\pm0.008$ (Vaughan et al. 2003). Using the scaling relations in Ponti et al. (2012), the black hole mass is estimated to be $2\times10^{6}~\rm M_{\odot}$. Jin, Done \& Ward (2016) revisited this relation using a more complete reverberation mapping sample and presented the regression results, from which we can derive a 2$\sigma$ range of 0.5-8.0 $\times10^{6}~\rm M_{\odot}$ for the black hole mass.

In addition, we can also obtain a rough estimate of the black hole mass from the scaling relation between the mass, bolometric luminosity and high frequency break in the X-ray PSD (M$^{c}$Hardy et al. 2007). Adopting a high frequency break of $6.4^{+4.7}_{-2.7}\times10^{-4}$ Hz as being observed in the 0.3-1 keV PSD (see Paper-I), and a bolometric luminosity of $4.2\times10^{45}$ erg s$^{-1}$ (see next section), the black hole mass is found to be $4.3^{+1.3}_{-1.0}\times10^{6}~\rm M_{\odot}$. Therefore, despite the low significance, the high frequency break in the 0.3-1 keV PSD also indicates a black hole mass within the 2$\sigma$ range from the X-ray variability method.

Considering all the above black hole mass estimates, as well as the mass of $1.8\times10^{7}~\rm M_{\odot}$ derived from the SED fitting in Section~\ref{sec-sed} which is most likely to be an over-estimate, a reasonable estimate of the black hole mass of \rxj04\ is likely to be $5-10~\times 10^{6}~\rm M_{\odot}$, which is slightly larger than the SMBH in \pg12\  (J13) and \1h07\ (DJ16).

\begin{table}
 \centering
   \caption{Comparison of the mass accretion rate through the outer disc ($\dot{m}_{\rm out}$) and the observed Eddington ratio ($L_{\rm bol}/L_{\rm Edd}$) for $L_{\rm 1}$ and $L_{\rm 2}$ in Fig.~\ref{fig-sed2}. We assume 30$\degr$ inclination angle and zero spin. A higher spin or a larger inclination angle will further increase the $\dot{m}_{\rm out}$ values (see DJ16).}
     \begin{tabular}{lcccc}
\hline
  BH Mass ($~\rm M_{\odot}$)& $5\times10^{6}$ & $7\times10^{6}$ & $1\times10^{7}$ & $1.8\times10^{7}$\\
  \hline
  $L_{1}/L_{\rm Edd}$ &6.5&4.6&3.2&1.8\\
  $L_{2}/L_{\rm Edd}$ &5.4&3.8&2.7&1.5\\
  $\dot{m}_{\rm out}$ &23.8&12.1&5.9&1.8\\
  \hline
   \end{tabular}
 \label{tab-mdot}
\end{table}

\subsection{Bolometric Luminosity and Mass Accretion Rate}
\label{sec-mdot}
The well-constrained SED of \rxj04, together with the low Galactic and intrinsic extinction, enables us to make one of the most reliable estimates of an AGN bolometric luminosity. Using $M=1.8\times10^{7}~\rm M_{\odot}$ and $L_{\rm bol}/L_{\rm Edd}=1.8$ based on the best-fit SED with zero spin, we can derive a bolometric luminosity of $L_{\rm bol}=4.2\times10^{45}$ erg s$^{-1}$ for a $30\degr$ inclination angle\footnote{The best-fit {\it maximal-spin} SED model gives $M=3.5\times10^{7}~\rm M_{\odot}$ and $L_{\rm bol}/L_{\rm Edd}=2.2$, which would then increase $L_{\rm 1}$ by a factor of 2.4. This is because the maximal-spin SED contains a stronger peak than the zero-spin SED in the unobservable far-UV band. But this black hole mass is too large compared to the other mass estimates in Section~\ref{sec-bhmass}, so the maximal-spin SED is not adopted in our further analysis.} ($L_{1}$ in Fig.~\ref{fig-sed2}). This SED model predicts a peak in emission between 100-150 \AA\ where no data exist, and so $L_{\rm 1}$ will contain some uncertainty. Therefore, we calculate a more conservative $L_{\rm bol}$ by simply linking the UV data to the soft X-ray data and integrating the luminosity below it (i.e. $L_{2}$ in Fig.~\ref{fig-sed2}). We find $L_{2}=3.5\times10^{45}$ erg s$^{-1}=0.83L_1$, which can be considered as a lower limit for the true $L_{\rm bol}$. $L_{\rm 1}$ can be considered as an upper limit because it does not consider any other forms of energy loss in the disc (see Section~\ref{sec-ad-energy}).

The well-constrained $L_{\rm bol}$ also allows us to accurately measure the bolometric corrections at 5100 \AA\ ($k_{5100}\equiv L_{\rm bol}/L_{5100}$) and the 2-10 keV band ($k_{2-10}\equiv L_{\rm bol}/L_{\rm 2-10~keV}$). We find $k_{5100}=$ 37 and 31, $k_{2-10}=$ 91 and 76 for $L_{\rm 1}$ and $L_{\rm 2}$, respectively. These values are significantly larger than the commonly adopted values (e.g. Kaspi et al. 2000: $k_{5100}=9$; Richards et al. 2006: $k_{5100}=10.3$), but are consistent with reported correlations between various bolometric corrections and Eddington ratio (Vasudevan \& Fabian 2007, 2009; Jin, Done \& Ward 2012c). The reason for this Eddington ratio dependence is that the broadband SED contains a much stronger big blue bump when the black hole mass is lower and the mass accretion rate is higher, thereby producing much larger bolometric corrections. Therefore, the results in this paper highlight the importance of using the Eddington ratio dependent bolometric corrections as given in e.g. Vasudevan \& Fabian (2009) and Jin, Done \& Ward (2012c), or alternatively by using the broadband SED to derive $L_{\rm bol}$ directly.

Now we can use $L_{\rm 1}$ and $L_{\rm 2}$ to estimate the Eddington ratio for different black hole masses using $L_{\rm bol}/L_{\rm Edd} \propto M^{-1}$ for a fixed $L_{\rm bol}$. We can also estimate the mass accretion rate through the outer disc (hereafter: $\dot{m}_{\rm out}$) using the optical/UV luminosity ($L_{\rm opt}$) and the relation $L_{\rm opt}\propto\ (M^2\dot{m}_{\rm out})^{2/3}\cos i$, where $i$ is the inclination angle (Davis \& Laor 2011; DJ16). Table~\ref{tab-mdot} compares $L_{\rm bol}/L_{\rm Edd}$ with $\dot{m}_{\rm out}$ for a range of black hole masses, and shows that $\dot{m}_{\rm out}$ is clearly much larger than $L_{\rm bol}/L_{\rm Edd}$ if the black hole mass is $\lesssim10^{7}~\rm M_{\odot}$. This strongly suggests that the accretion flow loses gravitational energy through physical processes other than radiation, the details of which are discussed in the next section.

\begin{figure}
\centering
\begin{tabular}{c}
\includegraphics[bb=65 200 558 648, scale=0.46]{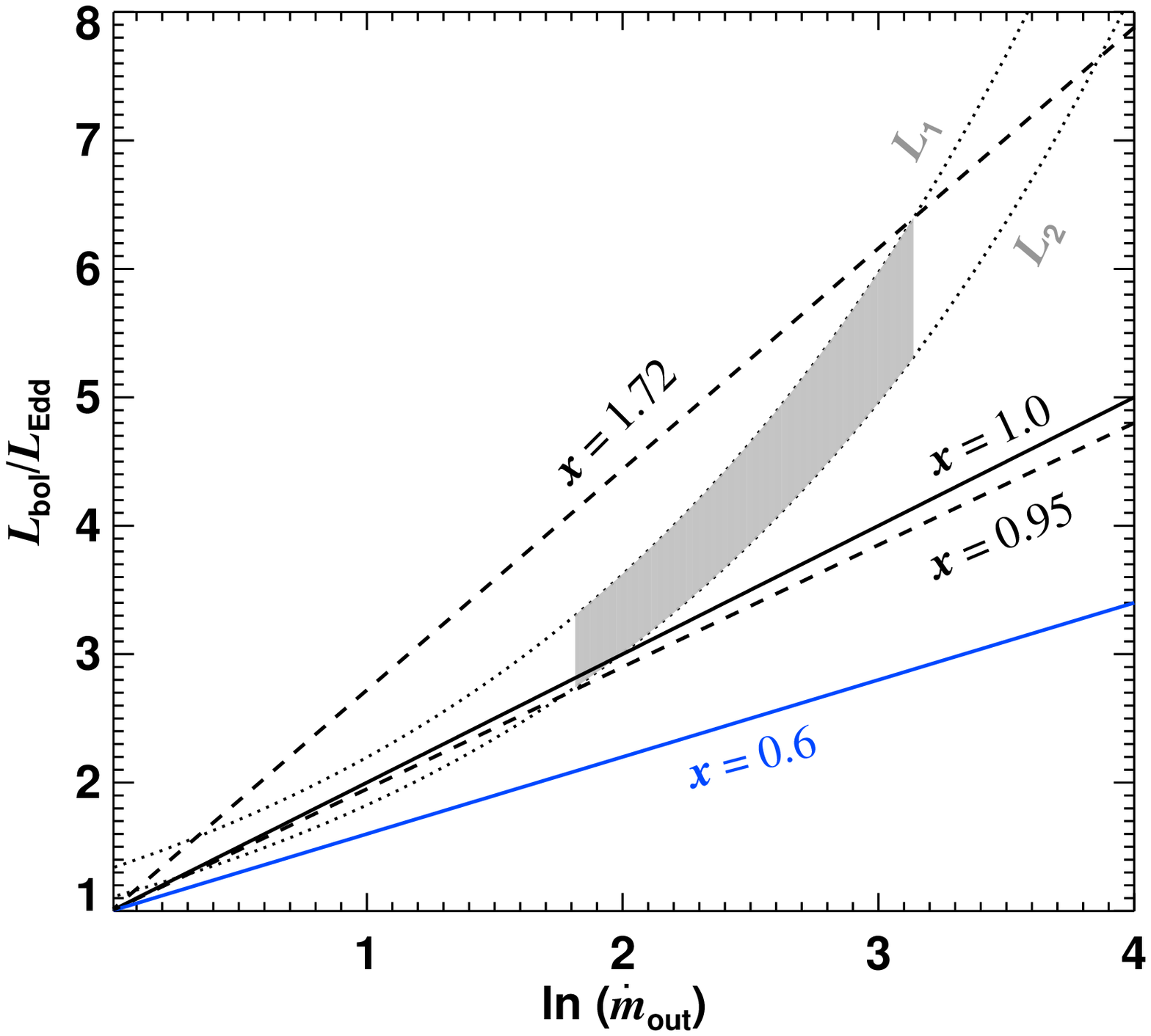} \\
\end{tabular}
\caption{Constraining the $\alpha$ factor in the relation $L_{\rm bol}/L_{\rm Edd} = 1+x\ {\rm ln}(\dot{m}_{\rm out})$. The two dotted curves indicates the $L_{\rm bol}/L_{\rm Edd}$ vs. ${\rm ln}(\dot{m}_{\rm out})$ relation for the observed optical/UV flux and the $L_{\rm bol}$ limits of $L_{\rm 1}$ and $L_{\rm 2}$ in Fig.~\ref{fig-sed2}. The grey region is defined by $L_2 \le L_{\rm bol} \le L_1$ and $5\times10^{6}~\rm M_{\odot} \le M \le 1.0\times10^{7}~\rm M_{\odot}$. $x=1$ (black solid line) and 0.6 (blue solid line) are predicted by the super-Eddington disc models considering only advection and outflow, respectively (see Poutanen et al. 2007). The two black dash lines indicate the range of $\alpha$ as constrained by the grey region.}
\label{fig-mdot}
\end{figure}

\subsection{Energy Loss through the Disc Wind and Advection}
\label{sec-ad-energy}
The SS73 standard accretion disc model can be applied when the gravitational energy of the disc material is fully thermalised and dissipated as radiation. In the case of a super-Eddington source ($\dot{m}_{\rm out}~>~1$) such as \rxj04,  the accretion flow is radiation pressure supported and is both geometrically and optically thick. In this case both advection and disc wind can carry away significant amount of disc energy, thereby reducing the energy radiated (e.g. SS73; Abramowicz et al. 1988; Lipunova 1999; Poutanen et al. 2007; Ohsuga \& Mineshige 2011; Takeuchi et al. 2014; Laor \& Davis 2014; Jiang, Stone \& Davis 2014; S\c{a}dowski \& Narayan 2015; Hashizume et al. 2015; Hagino et al. 2016; DJ16). In Fig.~\ref{fig-sed2}, the green curve shows a standard accretion disc spectrum with $M=7\times10^{6}~\rm M_{\odot}$ and $\dot{m}_{\rm out}=12.1$ in order to reproduce the observed optical/UV flux. This model clearly over-predicts the soft X-ray emission, with the largest discrepancy being a factor of 6.3. The Eddington ratio for $L_{\rm 1}$ is only 4.6  (i.e. 38\% $\dot{m}_{\rm out}$). Similar discrepancies have been found in other sources like \pg12, \1h07\ (DJ16), and is probably most severe in the intermediate mass black hole (IMBH) RX J1140.1+0307 (Jin, Done \& Ward 2016). All these sources have a $\dot{m}_{\rm out}$ that significantly exceeds the Eddington limit, suggesting that extra energy loss through a disc wind and/or advection exist in many super-Eddington sources.

Fig.~\ref{fig-sed2} shows that the discrepancy between a standard disc model and the data only exists in the far-UV and soft X-ray band, while Fig.~\ref{fig-uvspec} shows that the optical and near-UV spectrum is consistent the standard disc model. These results clearly indicate that a disc wind and/or advection only occurs within some critical radii, where the accretion flow starts to behave differently from a standard thin disc. Understanding the structure and physical properties of these extreme accretion flows requires detailed hydrodynamic simulations (e.g. Ohsuga et al. 2005; Okuda et al. 2005; Jiang, Stone \& Davis 2013; Jiang, Davis \& Stone 2016). Nevertheless we can gain a basic understanding of these flows by using simplified analytical calculations. In the case of a super-Eddington disc with advection (i.e. the so-called `slim' disc, Abramowicz et al. 1988), the critical radius ($R_{\rm crit}$) is where the photon escape time-scale is equal to the accretion time-scale. It has been found that $R_{\rm crit}\approx \dot{m}_{\rm out}R_{\rm in}$, where $R_{\rm in}$ is the inner radius, and $L_{\rm bol}/L_{\rm Edd} = 1+ {\rm ln}(\dot{m}_{\rm out})$ (Watarai et al. 2000; Pounanen et al. 2007). In the case of a super-Eddington disc with a disc wind, $R_{\rm crit}$ is where the half-thickness of the disc is equal to the radius (SS73; Bisnovatyi-Kogan \& Blinnikov 1977; Lipunova 1999), then $R_{\rm crit}\approx \dot{m}_{\rm out}R_{\rm in}$ and $L_{\rm bol}/L_{\rm Edd} = 1+0.6\ {\rm ln}(\dot{m}_{\rm out})$ (Pounanen et al. 2007). Numerical simulations show that both advection and disc wind can exist in a super-Eddington accretion flow (e.g. Eggum, Coroniti \& Katz 1988; Ohsuga et al. 2005; Okuda et al. 2005), in which case Pounanen et al. (2007) shows that $R_{\rm crit}$ has a weak dependance on the relative strength between the advection and the disc wind (their Equation 21).

Given its well constrained $L_{\rm bol}$ and $L_{\rm opt}$, \rxj04\ offers a great opportunity to test these theoretical results. The main uncertainty lies in the black hole mass. Since $L_{\rm bol}/L_{\rm Edd}$ and $\dot{m}_{\rm out}$ both depend on the black hole mass, we can derive a relation between these two parameters for $L_{\rm 1}$ and $L_{\rm 2}$ (i.e. the two dotted curves in Fig.~\ref{fig-mdot}). We can further constrain this relation by adopting a black hole mass range of $5-10~\times10^{6}~\rm M_{\odot}$ (the grey region in Fig.~\ref{fig-mdot}). For the relation $L_{\rm bol}/L_{\rm Edd} = 1+x~{\rm ln}(\dot{m}_{\rm out})$, the difference between advection and a disc wind lies in the $x$ factor, which is 1.0 for only advection and 0.6 for only disc wind. Fig.~\ref{fig-mdot} shows that $x=0.6$ under-estimates $L_{\rm bol}/L_{\rm Edd}$ for a specific $\dot{m}_{\rm out}$, while $x=1.0$ is roughly consistent with $L_{\rm bol}\sim L_2$ and $M\sim 10^{7}~\rm M_{\odot}$. The grey region constrains $x$ to be $0.95\le x \le1.72$, with a larger $x$ indicating a higher $L_{\rm bol}$ and a lower $M$. Therefore, our results suggest that the super-Eddington accretion flow in \rxj04\ tends to radiate more energy than predicted by the theoretical calculations for accretion disc models with advection and a disc wind.

In addition, assuming $R_{\rm in}=6~R_{\rm g}$ and using Equation 12 in Pounanen et al. (2007), $R_{\rm crit}$ can be calculated to be 140-170 $R_{\rm g}$, 70-82 $R_{\rm g}$ and 33-36 $R_{\rm g}$ for $M=5\times10^{6}~\rm M_{\odot}$, $7\times10^{6}~\rm M_{\odot}$ and $1\times10^{7}~\rm M_{\odot}$, respectively. These $R_{\rm crit}$ values all indicate that the SED flattening due to the disc wind and advection should emerge at $\lambda < 900$\AA, which is consistent with the SED shown in Fig.~\ref{fig-sed2}. However, we cannot further constrain $R_{\rm crit}$ from current observations due to the lack of data in the unobservable far-UV region.

\begin{figure*}
\centering
\begin{tabular}{cc}
\includegraphics[bb=65 216 666 720,scale=0.39]{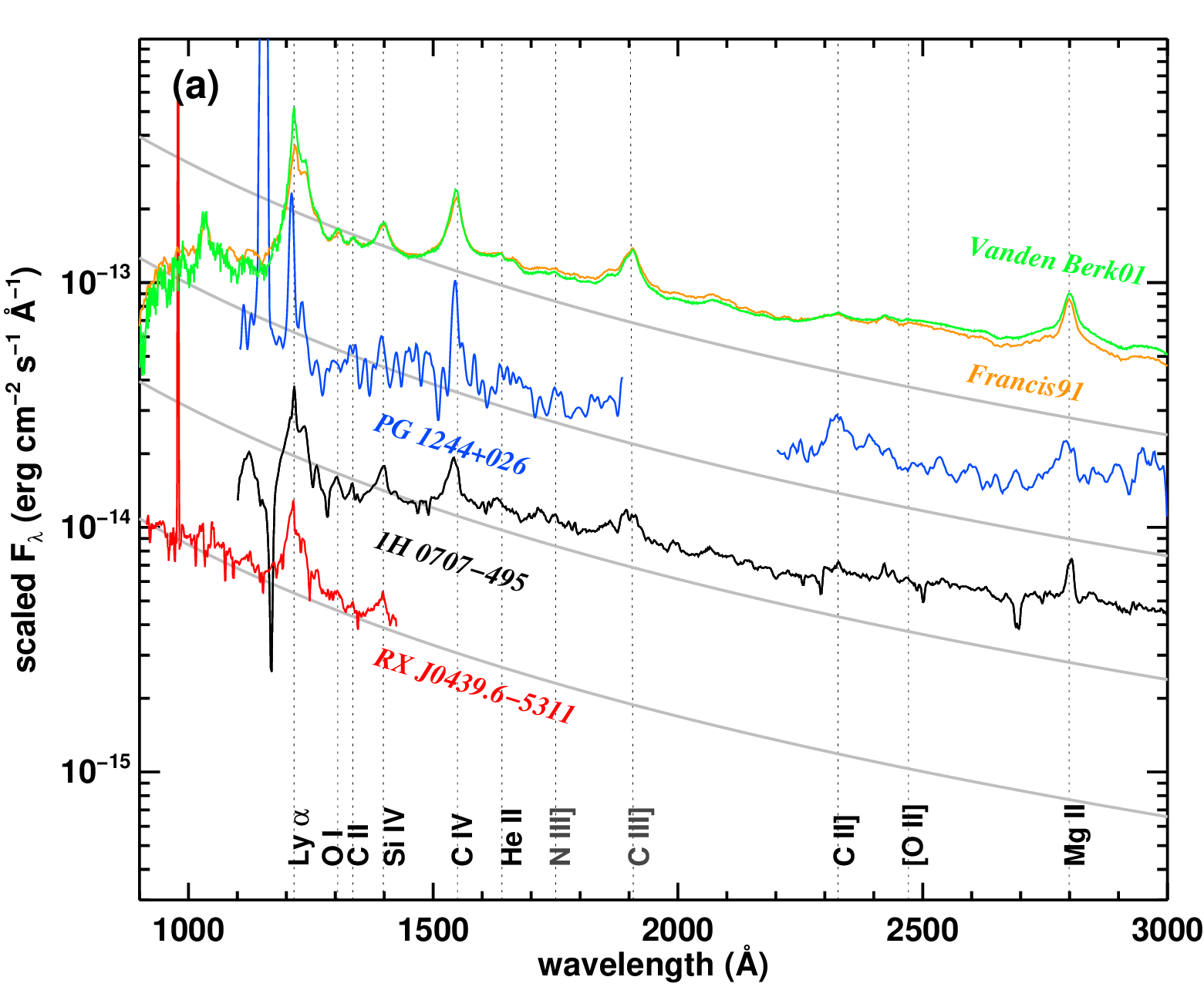} &
\includegraphics[bb=60 216 666 720,scale=0.39]{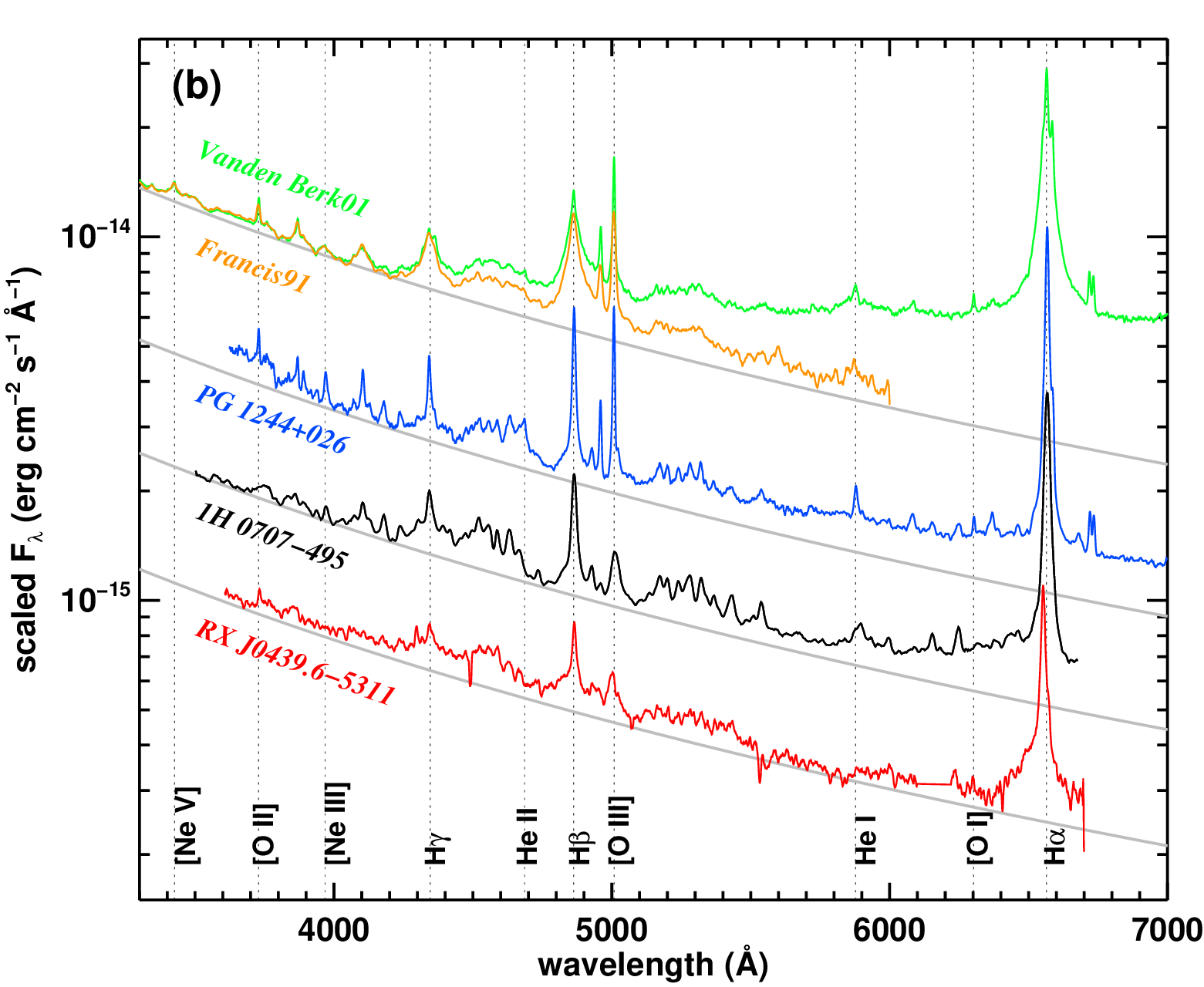} \\
\end{tabular}
\caption{Comparison of the rest-frame optical/UV continua and emission lines between \rxj04\ (red), \pg12\  (blue),\1h07\  (black), the bright quasar composite spectrum from Francis et al. (1991) (orange) and the SDSS quasar composite spectrum from Vanden Berk (2001) (green). Panel-a shows the comparison in the UV band, including the IUE spectra of \pg12\  and combined {\it HST} spectra of \rxj04\ and \1h07. Panel-b shows the comparison in the optical band, including the SDSS spectrum of \pg12, ESO optical spectrum of \rxj04\ and CTIO spectrum of \1h07\  (see DJ16). Each of the grey curves indicates the continual shape predicted by a standard thin disc rescaled to the flux level of each spectrum. Galactic reddening has been corrected for the three individual sources. The flux of \pg12\  and \1h07\  have been rescaled for clarity, while the flux of \rxj04\ is not changed.}
\label{fig-uvoptspec}
\end{figure*}

\begin{figure}
\centering
\begin{tabular}{c}
\includegraphics[bb=65 216 536 648, scale=0.5]{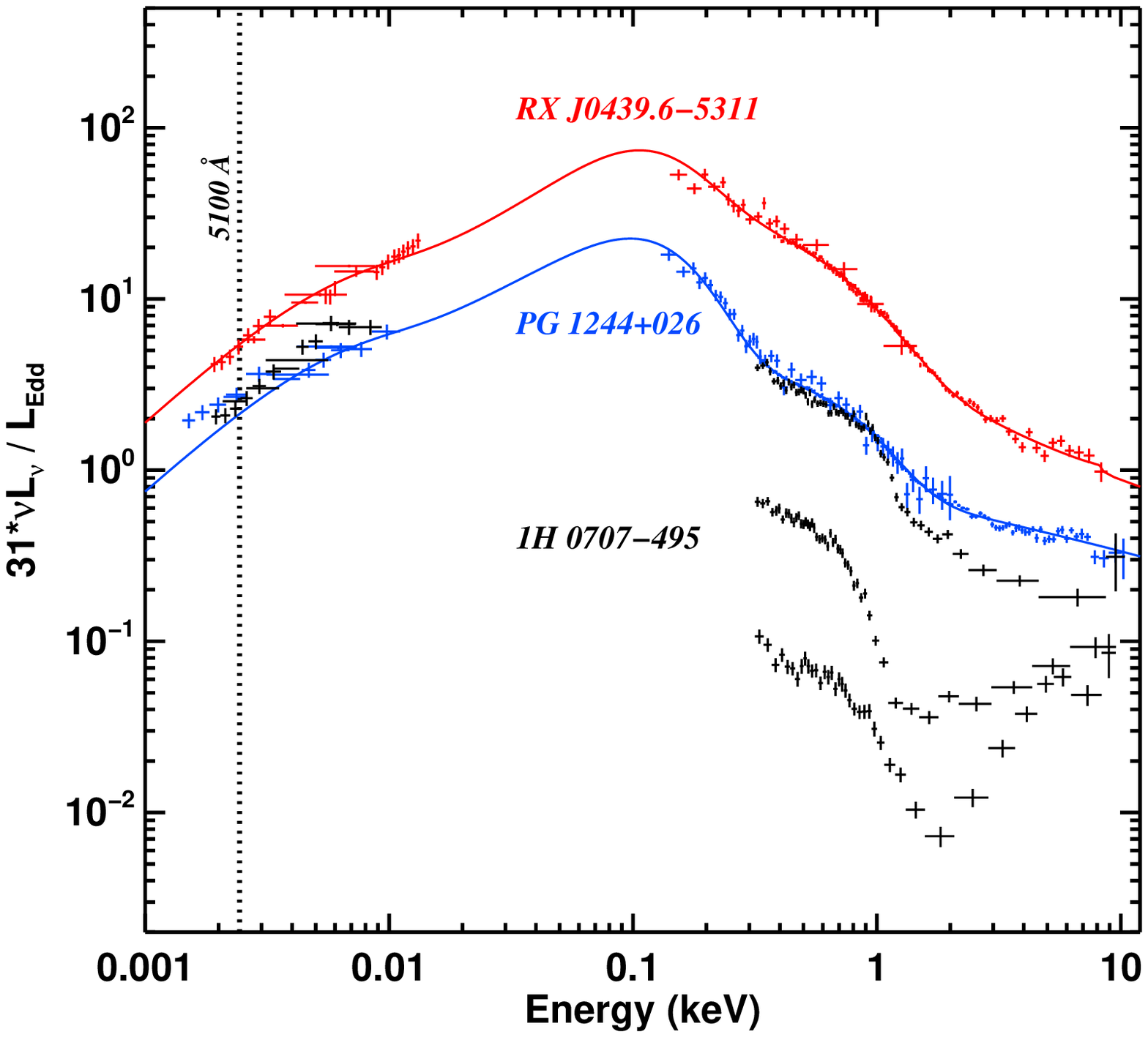} \\
\end{tabular}
\caption{Comparison of the rest-frame broadband SED between \rxj04\ (red), \pg12\  (blue) and \1h07\  (black). All the SEDs have been rescaled to show the Eddington ratio at 5100\AA\ with a bolometric correction of $k_{5100}=31$ (see Section~\ref{sec-mdot}). The three X-ray spectra of \1h07\  represent three typical spectral shape of this source (see DJ16). A host galaxy component is not included in the SED of \pg12\  (see J13), and is not required in the other two sources.}
\label{fig-sedcompare}
\end{figure}

\begin{table}
 \centering
   \caption{The distance of the soft X-ray excess ($R_{\rm SX}$) measured from the LF time lag ($\tau_{\rm LF}$) in \rxj04, PG 1244+026 (Jin, Done \& Ward 2013; Alston, Done \& Vaughan 2014), RE J1034+396 (Middleton et al. 2009; Kara et al. 2016) and RX J1140.1+0307 (Jin, Done \& Ward 2016). $\dot{m}_{\rm out}$ is the mass accretion rate measured from the optical/UV flux for the adopted black hole mass and zero spin. $R_{\rm hump}$ is the radius in a standard thin disc corresponding to $\sim1.8\times10^{5}K$, where an iron opacity hump can exist to produce a puffed-up disc region (Jiang, Davis \& Stone 2016) .}
     \begin{tabular}{@{}lccccc@{}}
\hline
Source & BH Mass & $\dot{m}_{\rm out}$ & $\tau_{\rm LF}$ & $R_{\rm SX}$ & $R_{\rm hump}$\\
& ($10^{6}~~\rm M_{\odot}$) &  & (ks) & ($R_{\rm g}$) & ($R_{\rm g}$) \\
\hline
  RX J0439 & 10 & 5.9 & $3.4\pm0.8$ & $70\pm16$ & 40 \\
  PG 1244 & 2 & 13 & $0.5\pm0.3$ & $50\pm30$ & 90 \\
  RE J1034 & 2 & 2 & $1.6\pm0.8$ & $160\pm80$ & 50\\
  RX J1140 & 1 & 10 & $0.6\pm0.3$ & $120\pm60$ & 110\\
\hline
   \end{tabular}
 \label{tab-rsx}
\end{table}

\subsection{Connection between the Soft X-ray Region and the Puffed-up Inner Disc Region}
A geometrically thick inner disc region in a super-Eddington accretion flow is predicted by theoretical calculations (e.g. Abramovicz et al. 1988; Wang \& Netzer 2003; Ohsuga \&Mineshige 2011), and supported by three-dimensional Magnetohydrodynamics (MHD) simulations of super-Eddington accretion discs (e.g. Jiang, Davis \& Stone 2014; S\c{a}dowski et al. 2014), and required by observations to explain the weak optical/UV emission lines (e.g. C {\sc iv}, He {\sc ii}, [O {\sc iii}] $\lambda$5007) in radio-quiet weak emission-line quasars (WLQs) which are accreting near/above their Eddington limits (e.g. Boroson \& Green 1992; Baskin \& Laor 2004; Shen \& Ho 2014; Shemmer \& Lieber 2015). Moreover, in WLQs it was also suggested that for the disc shielding mechanism to work efficiently to produce weak NLR and BLR lines, the puffed-up region has to be high enough and the X-ray corona region has to be compact ($\sim10~R_{\rm g}$, Luo et al. 2015 and references therein). However, these conditions are difficult to meet in NLS1s, especially for those with super-Eddington mass accretion rates and weak [O {\sc iii}] lines relative to Balmer lines (e.g. Jin et al. 2012a). This is because in sources like \rxj04\ and \pg12\ the distance from the SMBH to the soft X-ray region is at least tens of $R_{\rm g}$ (estimated from the light travel time), and the soft X-ray region radiates much more energy than the hard X-ray region. Furthermore, Jin, Ward \& Done (2012b) showed that [O {\sc iii}] $\lambda$5007 exhibits a much stronger correlation with the hard X-rays than with the soft X-rays. These results suggest that the NLR clouds can `see' the nuclear hard X-ray emission, but is shielded from the soft X-ray emission.

Recently, Jiang, Davis \& Stone (2016) applied their three-dimension MHD simulation of accretion disc around a SMBH to show that the iron opacity bump due to the bound-bound transition of Fe can increase the stability of the disc, change the disc structure and drive an outflow. Since this opacity bump only exists around $1.8\times10^{5}~K$ and is sensitive to the temperature, it only creates a puffed-up disc structure at a certain radial distance. For the `simple' NLS1s mentioned above, we can make a rough comparison between the radius of the soft X-ray region ($R_{\rm SX}$) and the radius of the $1.8\times10^{5}~K$ disc region ($R_{\rm hump}$). These radii depend on the black hole mass, mass accretion rate and the temperature structure of the super-Eddington accretion flow. Since each of these factors contains significant uncertainty, we can only expect to obtain an order-of-magnitude estimate. We use the time-lag between the soft and hard X-rays below $10^{-4}$ Hz to estimate the distance between the soft and hard X-ray corona (Paper-I), which is probably only a lower limit due to the multi-component dilution effect, while a more realistic measurement would require a full spectral-timing analysis which is then model-dependent (Gardner \& Done 2014). Then we use the temperature structure of a standard thin disc to estimate $R_{\rm hump}$, which is also likely to be a lower limit. This is because the effective disc temperature would be much higher for low mass and high mass accretion rate when the vertical disc structure is taken into consideration (e.g. Davis \& Hubeny 2006; Done \& Davis 2008; Done et al. 2012). However, since the outer part of a super-Eddington accretion disc is likely to remain as a standard thin disc (e.g. Davis \& Laor 2011), this estimate may still be roughly valid. In any case the uncertainty in black hole mass can easily introduce a factor of a few uncertainty to both $R_{\rm SX}$ and $R_{\rm hump}$.

Despite all potential uncertainties, we find that $R_{\rm SX}$ is roughly consistent with $R_{\rm hump}$ (Table~\ref{tab-rsx}), which indicates the possible connection between the soft X-ray region and the puffed-up disc region. The inner edge of the puffed-up disc region might be highly ionised and so could provide the low temperature, optically thick electrons required to produce the soft excess, as being described in Paper-I. This geometry naturally produces a large covering factor for the outer disc and emission line regions to be sufficiently shielded from the luminous soft X-ray region, without requiring the disc to be highly puffed-up or the X-ray emitting region to be very compact. Furthermore, in Paper-I we showed that the soft X-ray region provides seed photons for the hard X-ray corona. As $R_{\rm SX}$ increases with $\dot{m}_{\rm out}$, its covering factor for the hard X-ray corona may decrease, which then provides a possible explanation for the observed weaker hard X-ray emission in higher mass accretion sources (e.g. the strong correlation between $K_{2-10}$ and $L_{\rm bol}/L_{\rm Edd}$, Vasudevan \& Fabian 2007, 2009; Jin, Done \& Ward 2012c).

Finally, Paper-I showed that the soft excess in \rxj04\ is very smooth and so it does not favour any line features as would be present in the reflection spectrum modelled with {\sc rfxconv}. Even if the soft X-ray is modelled with the soft X-ray Comptonisation plus a weak disc reflection, a small $R_{\rm in}$ of 2.80$^{+2.74}_{-1.80}$ is still required to introduce sufficient relativistic smearing to the line features in the reflection component (see Figure 5 in Paper-I). However, we notice that the {\sc rfxconv} model is essentially based on the assumption of a constant-density atmosphere above the accretion disc (Ross \& Fabian 2005 and references therein), but the puffed-up disc region is more likely to be in the hydrostatic pressure equilibrium state. Done \& Nayakshin (2007) have shown that the soft excess in the hydrostatic models is much weaker than in the constant-density model, thus the reflection spectrum from the puffed-up disc region may be more smooth in the soft X-ray band, thereby relaxing the requirement for a very small $R_{\rm in}$ (or a very high black hole spin).

\subsection{A Unified Picture for the Accretion Flow in `Simple' and `Complex' Super-Eddington NLS1s}
\label{sec-unify}
DJ16 has shown that \pg12\ and \1h07\ have similar black hole masses
and mass accretion rates, and so their accretion flows may have
similar properties. Then the apparent differences in their
X-ray spectra can be explained as due to different viewing angles
relative to the clumpy disc wind. However, a remaining problem is that
their optical emission lines are very different from each other, with
\pg12\ having much stronger NLR lines than \1h07.
If the viewing angle scenario is correct,
there should be some NLS1s with similar optical/UV emission lines to
\1h07\ and similar broadband SED shape to \pg12. We find \rxj04\ is
indeed such a source.

Firstly, we compare these three NLS1s in terms of their optical/UV spectra. \pg12\ has IUE and SDSS spectra, while \1h07\ has {\it HST} STIS spectra and an optical spectrum from the Cerro Tololo Inter-American Observatory (CTIO). Since these spectra were not observed simultaneously, we perform the comparison in optical and UV band, separately. In addition, we include the bright quasar composite spectrum from Francis et al. (1991) and SDSS quasar composite spectrum from Vanden Berk (2001). In Fig.~\ref{fig-sedcompare}a,b we can clearly see that \rxj04\ and \1h07\ have very similar optical/UV spectra, implying similar intrinsic SEDs. Compared with the quasar composite spectra, these two sources have much weaker [O {\sc iii}]$\lambda$4959/5007 doublets whose ionisation potential is 55 eV, but some other lines of similar ionisation potentials such as C {\sc iv} (64 eV) and Si {\sc iv} (45 eV) have similar equivalent widths. Reverberation mapping studies have shown that C {\sc iv} and Si {\sc iv} emission line regions are much closer to the black hole than H$\beta$ and [O {\sc iii}] (e.g. Zu, Kochanek \& Peterson 2011; Peterson et al. 2014), thus these elements may still `see' the ionising source in the nuclear. Therefore, the optical/UV emission lines also depend on their location relative to the black hole and accretion disc. The NLR lines in \pg12\ are stronger than those seen in the quasar composite spectra which are dominated by the AGN emission, and so is consistent with the possibility that some of the narrow line emission in \pg12\ may come from an extended NLR region (ENLR) where ionisation by young massive stars in the host galaxy is likely to dominate (Unger et al. 1987; Husemann et al. 2014).

We notice that the spectra of \pg12\ and \1h07\ and the two quasar composite spectra all appear flatter than the standard thin disc model. Since we only corrected for the Galactic reddening for the three NLS1s, part of this flatness might be due to some intrinsic reddening, especially for the two composite spectra. The presence of some contribution from a host galaxy can also produce a flattening in the optical-UV continuum, which is most likely in \pg12\ and \1h07\ whose host galaxies can be resolved. However, \rxj04\ is an AGN at z=0.242 with no detectable intrinsic reddening, and so it is not affected by the host galaxy star-light or reddening. This is also consistent with the fact that its optical/UV continuum is most consistent with a standard disc. Reprocessing occurring within the disc could also contribute to the continuum flattening, but it is difficult to quantify without simultaneous optical and UV spectra.

We also compare the broadband SED from optical to hard X-ray bands among these three NLS1s. DJ16 conducted a detailed comparison of the broadband SED between \pg12\ and \1h07. They reported that for the same black hole mass of $2\times10^{6}~\rm M_{\odot}$, 30$\degr$ inclination angle and zero spin, both sources have $\dot{m}_{\rm out}\gtrsim10$. \1h07\ is $\sim$50\% more luminous than \pg12\ in the optical/UV band, but is always fainter than \pg12\ in the X-ray band with two orders of magnitude variability. Now we add \rxj04\ to this comparison by rescaling the SED by a factor of $k_{5100}=31$ (Section~\ref{sec-mdot}), so that the y-axis value at 5100\AA\ directly indicates the Eddington ratio. Fig.~\ref{fig-sedcompare} shows that \rxj04\ has a very similar broadband SED to \pg12, except that it is a factor of 3 more super-Eddington.

\begin{figure*}
\includegraphics[bb=430 -20 712 520, scale=0.4]{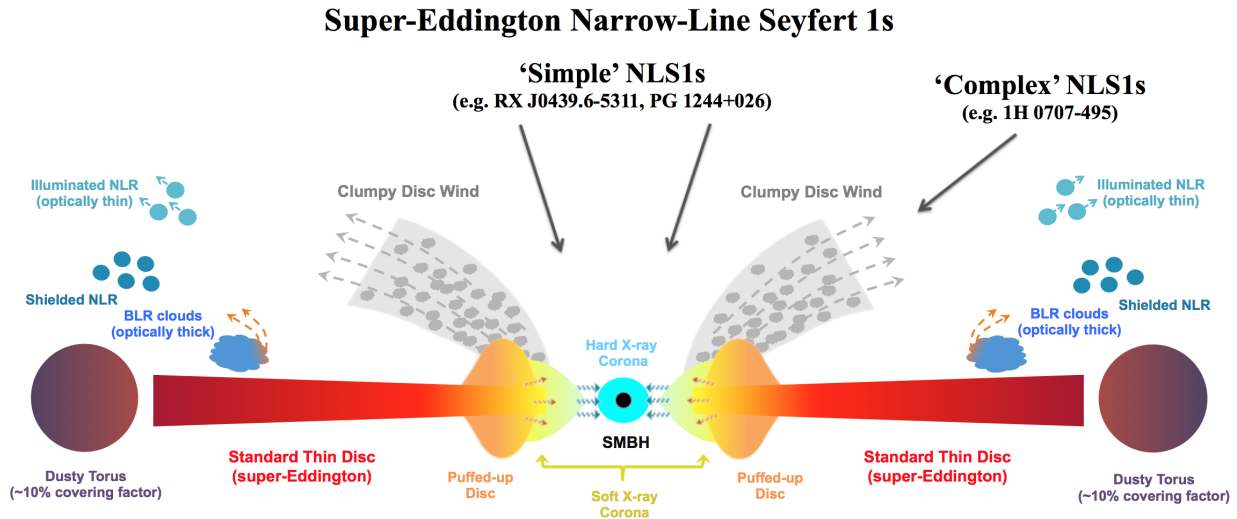}
\caption{A schematic cartoon of the super-Eddington accretion flow in \rxj04\ based on the results of our multi-wavelength study. We propose that this accretion flow picture may also be applicable for super-Eddington NLS1s such as \pg12, 1 H0707-495, Mrk 335, etc, with the apparent differences in their X-ray emission caused by the viewing angle effect relative to the clumpy disc wind (see Section~\ref{sec-ad-energy} for detailed descriptions related to this picture).}
\label{fig-cartoon1}
\end{figure*}

Summarising all the above comparison, we find that \rxj04\ has similar optical/UV spectra to \1h07\ and similar SED to \pg12, thereby providing good evidence that `simple' and `complex' super-Eddington NLS1s can indeed be unified in the inclination angle scenario as shown in Fig.~\ref{fig-cartoon1}. Both \rxj04\ and \pg12\ have low inclination angles with clear line-of-sights directly to the core region. But \1h07\ has a larger inclination angle, and so the disc wind material can intervene in the line-of-sight to the nuclear region, thereby absorbing X-rays and introducing extra X-ray variability.

\subsection{Super-Eddington NLS1s as the Low-redshift Analogs of Weak Emission-line Quasars}
\label{sec-wlq}
We compare the super-Eddington NLS1s with the radio-quiet WLQs at high redshifts. WLQs are well-known for their weak UV/optical high ionisation lines and prominent UV Fe {\sc ii} and Fe {\sc iii} emission (e.g. Diamond-Stanic et al. 2009; Plotkin et al. 2010a,b; Wu et al. 2012; Luo et al. 2015 and references therein). PHL 1811 is one of the most extreme and best studied WLQ, whose optical/UV spectra show little forbidden or semi-forbidden lines (e.g. Leighly et al. 2007). It has been suggested that WLQs can be divided into two subtypes according to their X-ray luminosity, about half of WLQs are X-ray normal, while the other half are X-ray weak (including PHL 1811 and its analogs, e.g. Shemmer et al. 2009; Wu et al. 2012; Luo et al. 2015). The spectral stacking results in Luo et al. (2015) indicate that the X-ray weakness of WLQs is mainly due to the absorption rather than a result of them being intrinsically X-ray weak. Then a shielding-gas scenario was proposed to unify these two WLQ subtypes. In this picture WLQs are all intrinsically similar, with X-ray weak WLQs being observed at larger inclination angles through a geometrically thick inner disc region (or a puffed-up/slim inner disc region), which blocks the line-of-sight to the nuclear X-ray emission (e.g. Wu et al. 2011; Luo et al. 2015). The puffed-up region may also act as a screen which shields the BLR and NLR from the nuclear ionising continuum, resulting in much weaker UV/optical emission lines. The high mass accretion rate of WLQs ($L_{\rm bol}/L_{\rm Edd}\gtrsim1$) also supports the existence of such a puffed-up inner disc region (e.g. Abramowicz et al. 1988; Ohsuga \& Mineshige 2011; Netzer \& Trakhtenbrot 2014).

We notice that the unified picture of super-Eddington NLS1s presented in this work is similar to that suggested for WLQs. The differences between `simple' and `complex' NLS1s are quite similar to the differences between the two WLQ subtypes. The two NLS1 subtypes have similar intrinsic optical/UV spectra (except the host galaxy emission), but the X-ray emission of `complex' NLS1s can be much fainter (Fig.~\ref{fig-sedcompare}). These NLS1s and WLQs are all accreting near/above the Eddington limit, and they all show intrinsically weak forbidden lines such as [O {\sc iii}]$\lambda5007$. Therefore, from one aspect the unified scenario of WLQs based on the inclination angle also support a similar unified scenario for super-Eddington NLS1s.

However, UV emission lines in NLS1s and bright quasar composite spectra are much stronger than those in WLQs, in particular the C {\sc iv} and Si {\sc iv} lines. This implies that the shielding material in NLS1s must be located at a larger radius than that of the C {\sc iv} and Si {\sc iv} emitting region. Since NLS1s typically have black hole masses of 1-3 orders of magnitude smaller than quasars, their $\dot{m}_{\rm out}$ can be much more super-Eddington and their disc can be much hotter, so the radius where the disc becomes geometrically thick can be much larger as well. Luo et al. (2015) showed that the puffed-up disc radius in WLQs is a few tens of $R_{\rm g}$, while we showed that for \rxj04\ the $R_{\rm crit}$ can be hundreds of $R_{\rm g}$ (Section~\ref{sec-ad-energy}). Therefore, the difference in the UV emission lines between NLS1s and WLQs can be explained by different relative locations and sizes of their puffed-up inner disc regions.

In effect we could consider these super-Eddington NLS1s as the low redshift analogs of WLQs, with `simple' NLS1s corresponding to X-ray normal WLQs, and `complex' NLS1s corresponding to X-ray weak WLQs. However, there are still some questions that remain to be answered. For example, NLS1s are likely to have higher Eddington ratios than WLQs, and  a disc wind may play a more critical role. Indeed, Gardner \& Done (2015) and Hagino et al. (2016) have shown that the obscuration by clumps in the disc wind can reproduce the observed X-ray light curves and spectra of \1h07. It is not clear whether or not the presence of a disc wind plays a significant role in the X-ray weakness of some WLQs. Additional X-ray variability is expected if there is strong wind absorption in an X-ray weak WLQ, but the timescale of this variability may scale up as the black hole mass, thereby making the variability difficult to detect in a single observation. We cannot find any study on the long-term X-ray variability in any X-ray weak WLQs. Another question is that whether the disc wind in NLS1s could also be responsible for their weak Oxygen forbidden lines or not. Moreover, if the Eddington ratio is the only key parameter required to explain the weak optical/UV lines in WLQs, then we should expect the same phenomena in all super-Eddington quasars. But this is clearly not the case in PG 1247+267, which is a non-WLQ at $z=2.038$ with well measured $L_{\rm bol}/L_{\rm Edd}=11$ (Trevese et al. 2014; Bentz \& Katz 2015; Lanzuisi et al. 2016). So there must be some other important parameters affecting the optical/UV line intensity. Future studies of these super-Eddington AGN are necessary in order to obtain a deeper understanding of these most extreme accretion flows in the universe.

\section{Summary and Conclusions}
In this paper we report the results from one of the most detailed multi-wavelength studies of an unobscured, highly super-Eddington QSO \rxj04. Firstly we found a better redshift of 0.242 for this source. Then the excellent multi-wavelength data-set enables us to make confident estimates of its $L_{\rm bol}$ ($\sim$20\%  uncertainty). Our results clearly indicate that it is one of the most robust `super-Eddington' source. For a black hole mass of $(5-10)\times10^{6}~\rm M_{\odot}$, we measure $L_{\rm bol}/L_{\rm Edd} = 2.7-6.5$ and $\dot{m}_{\rm out}=5.9-23.8$.

With the above $M$, $L_{\rm bol}$ and $\dot{m}_{\rm out}$, the multi-wavelength properties of the super-Eddington accretion flow in \rxj04\ can be summarised in the unified picture in Fig.~\ref{fig-cartoon1}. The key aspects of this picture include the following:
\begin{itemize}
\item the outer part of the accretion flow is consistent with the standard thin disc model outside of 190-380 $R_{\rm g}$ for a black hole mass of $5-10~\times10^{6}~\rm M_{\odot}$. This is suggested by the underlying continua in the optical/UV spectra observed down to 900 \AA.
\item within a critical radius (e.g. $140-170~R_{\rm g}$ for $M=1\times10^{7}~\rm M_{\odot}$ and $\dot{m}_{\rm out}=5.9$), the accretion disc starts to deviate from a standard thin disc model because of a strong disc wind and/or advection, as could be expected in such high mass accretion rate flows, which carries away a significant amount of the gravitational energy in the disc. This is supported by the large difference between $L_{\rm bol}/L_{\rm Edd}$ and $\dot{m}_{\rm out}$ in Table~\ref{tab-mdot}, as well as the SED discrepancy in the far-UV and soft X-ray band shown in Fig.~\ref{fig-sed2}.
\item at some radii of at least tens of $R_{\rm g}$ away from the black hole, there exists a low temperature thermal electron population, which up-scatters disc photons into the soft X-ray band to produce the prominent soft excess. Then some of these soft X-ray photons meet the hot corona at small radii and are up-scattered into the hard X-ray band. This inner disc structure is supported by the results of X-ray spectral-timing study in Paper-I. We also find tentative evidence for the connection between the soft X-ray region and the puffed-up inner disc region.
\item The geometrically thick inner disc and/or the clumpy disc wind can shield part of NLR from the nuclear ionising continuum, thereby reducing the strength of forbidden lines in the NLR such as [O {\sc iii}]. The extremely powerful radiation may also trigger a global outflow in the [O {\sc iii}] emitting NLR. The disc radiation may also be strong enough to trigger an outflow on the surface of the optically thick BLR clouds. Evidence for this is found in the line profiles of [O {\sc iii}]$\lambda$5007 and H$\beta$ (see Fig.~\ref{fig-hbfit}).
\item the hot dust in the torus has a covering factor of 10.8\% and produces essentially all of the near-IR emission (see Fig.~\ref{fig-sed2}).
\end{itemize}

\rxj04\ also allows us to constrain the $x$ factor in the $L_{\rm bol}/L_{\rm Edd}=1+x\ ln~\dot{m}_{\rm out}$ relation to be $0.95\le x \le 1.72$. Using a simulation based on the standard accretion disc model, we can rule out the possibility of detecting any reprocessed X-ray variability in the optical/UV disc emission, but a significant long-term optical/UV/X-ray covariance has been observed, which is likely caused by the long-term fluctuation of $\dot{m}_{\rm out}$.

Furthermore, we compare \rxj04\ with two other super-Eddington NLS1s \pg12\ and \1h07. We found that \rxj04\ behaves as a `simple' NLS1 in terms of its X-ray emission and broadband SED, but it is more similar to a `complex' NLS1 such as \1h07 in the optical/UV spectra. Therefore, the scenario we proposed for the super-Eddington accretion flow in \rxj04\ is likely to be applicable for both `simple' and `complex' super-Eddington NLS1 subtypes. Then different inclination angles lead to different line-of-sights to the nuclear X-ray emitting region, which then result in the observed differences in their X-ray spectra due to the possible passage of the line-of-sight through the clumpy wind material (see Fig.~\ref{fig-cartoon1}). Currently, a sample of representative NLS1s of high mass accretion rates is under our investigation in order to further verify this unified scenario. The results will be reported in the near future.

Finally, we also compare super-Eddington NLS1s with WLQs, finding that these two AGN populations share a range of similar properties from optical to hard X-rays, and that the inclination angle plays an crucial role in both of them. Therefore, we propose that the super-Eddington NLS1s could be the low-redshift analogs of WLQs at high redshift. However, there are also some obvious spectral variability differences between these extreme NLS1s and WLQs, which are probably caused by the large difference in their black hole masses and spins. Future work is required to compare these two AGN populations in more detail.

\section*{Acknowledgements}
We thank Dirk Grupe for kindly providing the unique optical spectrum of \rxj04\ observed with the ESO's 1.52 m telescope. The anonymous referee is appreciated for providing valuable comments and suggestions to improve the paper. CJ acknowledges the support by the Bundesministerium f\"{u}r Wirtschaft und Technologie/Deutsches Zentrum f\"{u}r Luft- und Raumfahrt (BMWI/DLR, FKZ 50 OR 1408 and FKZ 50 OR 1604) and the Max Planck Society. CD and MJW acknowledge STFC funding under grant ST/L00075X/1.

This work is mainly based on a recent observation conducted by \xmm, an ESA
science mission with instruments and contributions directly funded by
ESA Member States and the USA (NASA). We have
made use of the {\it ROSAT} Data Archive of the Max-Planck-Institut f\"{u}r
extraterrestrische Physik (MPE) at Garching, Germany.
We acknowledge the use of public data from the {\it Swift} data archive.
This work also makes use of observations conducted by the NASA/ESA Hubble Space Telescope, with public data obtained from the Mikulski Archive for Space Telescopes (MAST). STScI is operated by the Association of Universities for Research in Astronomy, Inc., under NASA contract NAS5-26555. Support for MAST for non-{\it HST} data is provided by the NASA Office of Space Science via grant NNX09AF08G and by other grants and contracts.
This work involves data products from the Wide-field Infrared Survey Explorer, which is a joint project of the University of California, Los Angeles, and the Jet Propulsion Laboratory/California Institute of Technology, funded by the National Aeronautics and Space Administration. We use data products from the Two Micron All Sky Survey, which is a joint project of the University of Massachusetts and the Infrared Processing and Analysis Center/California Institute of Technology, funded by the National Aeronautics and Space Administration and the National Science Foundation.
This research has made use of the NASA/IPAC Extragalactic Database (NED) which is operated by the Jet Propulsion Laboratory, California Institute of Technology, under contract with the National Aeronautics and Space Administration.











\bsp	
\label{lastpage}
\end{document}